\documentclass[aps,pra,twocolumn,superscriptaddress,showpacs]{revtex4-2}

\usepackage{amsmath,amssymb}
\usepackage{braket}
\usepackage{booktabs}
\usepackage{graphicx}
\usepackage{bm}
\usepackage{multirow}
\usepackage{cleveref}
\usepackage{float}
\usepackage{physics}
\usepackage{xcolor}

\makeatletter
\newcommand{\equalcontrib}{%
  \frontmatter@footnote{These authors contributed equally to this work.}%
}
\makeatother

\begin{document}

\title{Improving the Loss Tolerance of Heralded Photonic GHZ States for Long-Distance Device-Independent Conference Key Agreement}
% \authormark{*}
\author{Yazeed K. Alwehaibi \equalcontrib}
\email{y.alwehaibi23@imperial.ac.uk}
\affiliation{Blackett Laboratory, Department of Physics, Imperial College London, Prince Consort Rd, London, SW7 2AZ, United Kingdom}

\author{Makoto Ishihara \equalcontrib}
\email{llmakomako.arg1076@keio.jp}
\affiliation{Department of Electronics and Electrical Engineering, Keio University, 3-14-1 Hiyoshi, Kohoku-ku, Yokohama 223-8522, Japan}

\author{Shakib Daryanoosh}
\affiliation{Curtin Centre for Optimisation and Decision Science, Curtin University, Whadjuk Country, Perth 6102, Australia}

\author{Ewan Mer}
\affiliation{Blackett Laboratory, Department of Physics, Imperial College London, Prince Consort Rd, London, SW7 2AZ, United Kingdom}

\author{Shang Yu}
\affiliation{Blackett Laboratory, Department of Physics, Imperial College London, Prince Consort Rd, London, SW7 2AZ, United Kingdom}

\author{Wojciech Roga}
\affiliation{Department of Electronics and Electrical Engineering, Keio University, 3-14-1 Hiyoshi, Kohoku-ku, Yokohama 223-8522, Japan}
\author{Ian A. Walmsley}
\affiliation{Clarendon Laboratory, University of Oxford, Parks Road, Oxford OX1 3PU, UK}

\author{Masahiro Takeoka}
\email{takeoka@elec.keio.ac.jp}
\affiliation{Department of Electronics and Electrical Engineering, Keio University, 3-14-1 Hiyoshi, Kohoku-ku, Yokohama 223-8522, Japan}
\affiliation{
National Institute of Information and Communications Technology (NICT), Koganei, Tokyo 184-8795, Japan.
}

\author{Raj B. Patel}
\email{raj.patel1@imperial.ac.uk}
\affiliation{Blackett Laboratory, Department of Physics, Imperial College London, Prince Consort Rd, London, SW7 2AZ, United Kingdom}

\date{\today}

%==================

\begin{abstract}
Heralded multipartite entanglement distribution is a key requirement for device-independent conference key agreement (DI-CKA) over lossy quantum networks. Although locally equivalent in the absence of loss, different single-rail photon-number encodings of Greenberger--Horne--Zeilinger (GHZ) states respond differently to photon loss. Here, we investigate the critical detection efficiencies for detection-loophole-free parity--CHSH violations of computational-basis GHZ states---a coherent superposition of the vacuum and an $n$-photon component---and of fixed-photon-number GHZ states, deriving exact analytical conditions for both. We show that for states that are not permutation symmetric, such as the latter, the assignment of measurement roles to physical modes affects loss tolerance. Building on this, we introduce a star-network protocol employing heterogeneous sources to directly herald the loss-tolerant vacuum–$n$-photon GHZ states while retaining the favourable long-distance scaling $O(\eta_{\text{c}}^{n/2})$, where $\eta_{\text{c}}$ is the channel transmittance. For four users, we characterize the heralded state under photon loss and show that tunable source parameters allow genuine multipartite entanglement to persist at any finite channel distance. With ideal Pauli and displacement-based measurements, our protocol achieves positive DI-CKA key rates at lower detection efficiencies than previous schemes, while retaining comparable or greater rates and communication distances at high efficiency. Overall, our work improves the loss tolerance of heralded photonic GHZ states for DI-CKA both by directly heralding a more loss-tolerant encoding and by optimizing existing schemes. These results identify photon-number encoding, source architecture, and measurement-role assignment as key design parameters for loss-resilient multipartite quantum networks, offering a practical route toward near-term DI-CKA over long distances.
\end{abstract}

\maketitle

\section{Introduction}

Multipartite entanglement, exemplified by Greenberger–Horne–Zeilinger (GHZ) states \cite{greenberger1989going,greenberger1990bell}, provides a paradigmatic setting for probing the incompatibility of quantum mechanics with local realism \cite{mermin1990extreme,brunner2014bell}. Beyond its foundational significance, multipartite entanglement is a central resource for quantum networks, enabling applications such as conference key agreement \cite{epping2017multi,proietti2021experimental}, secret sharing \cite{hillery1999quantum,zhou2018quantum}, distributed sensing \cite{eldredge2018optimal,xia2020demonstration,zaporski2025quantum, zang2026quantum}, and networked quantum computation \cite{cirac1999distributed,nickerson2014freely}. Of particular importance is the generalized $n$-partite GHZ state,
\begin{equation}
    \ket{\Phi_n(q)}
    =\sqrt{1-q}\;\ket{0}^{\otimes n}\pm\sqrt{q}\;\ket{1}^{\otimes n},
    \label{eq:phi_n}
\end{equation}
where $0<q<1$. Its multipartite correlations can be used to violate Bell inequalities and certify security without relying on a detailed characterization of the used devices. These properties make GHZ states natural resources for device-independent conference key agreement (DI-CKA), in which $n$ spatially separated parties establish a common secret key from observed nonlocal correlations \cite{ribeiro2018fully, holz2020genuine, murta2020quantum}.

The practical performance of DI-CKA is, however, severely limited by optical loss. In a direct-transmission architecture, an $n$-partite photonic GHZ state is generated at a central station and one subsystem is transmitted to each user \cite{proietti2021experimental}. If every transmitted subsystem arrives successfully, the distribution probability scales as $O(\eta_{\text{c}}^n)$, where $\eta_{\text{c}}$ denotes the effective individual link transmittance, into which the heralding detector efficiency can also be absorbed. Local loss and detector inefficiency at the user nodes further suppress the observed Bell violation and can open the detection loophole, thereby precluding device-independent security. These effects become increasingly restrictive as either the distance or the number of users increases.
%=====
Heralded entanglement distribution provides a route around the first of these limitations. Rather than transmitting a pre-generated multipartite state, each user prepares a local two-mode entangled state, retains one mode, and sends the other to a central station. With an appropriately designed interferometer and conditional photodetection, a successful heralding event requires only $n/2$ transmitted photons to be detected for even $n$. The resulting distribution rate therefore scales as $O(\eta_{\text{c}}^{n/2})$, rather than the $O(\eta_{\text{c}}^n)$ scaling of direct transmission, and can substantially outperform direct distribution over lossy links as proposed in \cite{shimizu2025simple} for GHZ states or in \cite{roga2023efficient} for W and Dicke states. This improvement is reminiscent of the transmittance square-root advantage underlying the twin-field quantum key distribution, generalized to multipartite entanglement distribution \cite{lucamarini2018overcoming}.

Recent protocols have used this approach to distribute fixed-photon-number GHZ-like states of the form \cite{shimizu2025simple,ishihara2025longCKA,singh2026fault}
\begin{equation}
    \ket{\Psi_n(q)}
    =\sqrt{1-q}\;\ket{01}^{\otimes n/2}\pm\sqrt{q}\;\ket{10}^{\otimes n/2}.
    \label{eq:psi_n}
\end{equation}
Ishihara \emph{et al.}\ recently incorporated this heralding scheme into a DI-CKA protocol, computing key rates based on the parity--Clauser-Horne-Shimony-Holt (CHSH) inequality under two measurement scenarios, and demonstrating that efficient heralding can substantially extend the achievable communication distance relative to direct transmission \cite{ishihara2025longCKA}.

%======
As abstract qubit states, the \textit{vacuum-$n$-photon} state ($\ket{\Phi_n}$) is related to the \textit{fixed-photon-number} state ($\ket{\Psi_n}$) by local bit flips and is therefore locally-unitarily equivalent to it. The two states are equally entangled and give identical Bell violations in the lossless case. This equivalence does not, however, imply identical operational performance in single-rail photon-number encoding. In this setting, a bit flip exchanges vacuum and one photon and is therefore not a passive photon-number-conserving operation. Moreover, the two GHZ encodings respond differently to photon loss, and this distinguishes sharply between the two. Both branches of $\ket{\Psi_n}$ contain photons and are therefore affected by loss, whereas $\ket{\Phi_n}$ contains a vacuum branch that is invariant under loss. Thus, the relevant distinction is not entanglement, but rather the photon-number structure. In the bipartite case, this asymmetry is what allows $\ket{\Phi_2}$ to reach the Eberhard detection-efficiency limit, while $\ket{\Psi_2}$ does not \cite{alwehaibi2025tractable,ishihara2025QKD}. 
This naturally raises two questions: Does the same advantage persist in the multipartite regime, where it would be relevant to DI-CKA? If so, can the state $\ket{\Phi_n(q)}$ be efficiently heralded over lossy channels while retaining the favourable $O(\eta_{\text{c}}^{n/2})$ scaling?

The multipartite setting adds a further consideration to this comparison. The parity--CHSH test does not treat all parties equally. Two parties each choose between two measurement settings, while the remaining $n-2$ parties perform a single fixed measurement. Which physical modes are assigned these two-setting roles is therefore a free choice of the protocol. For the permutation-symmetric state $\ket{\Phi_n}$, this choice is immaterial. For $\ket{\Psi_n}$, however, the modes differ in their photon occupation within each component, and the loss tolerance consequently depends on how the measurement roles are assigned.

In this work, we derive exact detection-efficiency thresholds for parity--CHSH violations by both GHZ encodings. Under the conventional assignment of the two-setting roles, $\ket{\Phi_n(q)}$ offers greater loss tolerance than $\ket{\Psi_n(q)}$, with the difference most pronounced in the few-user regime relevant to near-term DI-CKA. We then show that assigning these roles to parties with equal photon occupations in $\ket{\Psi_n(q)}$ lowers its Bell-violation threshold to that of $\ket{\Phi_n(q)}$. Nevertheless, the two encodings can retain different key-generation thresholds because conference key agreement depends on correlations with every user.

Motivated by this advantage, we devise a star-network protocol that directly heralds the canonical GHZ state $\ket{\Phi_n(q)}$ over lossy channels while retaining the favourable $O(\eta_{\text{c}}^{n/2})$ long-distance scaling. We achieve this using a \emph{heterogeneous-source} architecture in which different users prepare different classes of initial states. Such direct heralding cannot be achieved using an otherwise equivalent network of identical sources \cite{shimizu2025simple,ishihara2025longCKA}.

We characterize the protocol in the presence of channel loss and relevant experimental imperfections. For four users, we derive the conditional output state, heralding probability, and fidelity with the target computational-basis GHZ state, and compare its performance with direct transmission and with heralded distribution based on the previously considered GHZ-like encoding. We additionally show that, by tuning the parameters of the initial states, genuine multipartite entanglement can be certified at any finite channel distance.

Combining these results, we compute the DI-CKA key rate for $n=4$ as a function of distance and detection efficiency under two measurement scenarios: ideal Pauli measurements in the $XZ$ plane and the more experimentally relevant displacement-based measurements. We compare the achievable key rates of our protocol with those of the previously proposed identical-source, fixed-photon-number scheme, allowing measurement roles to be optimized for the latter, and with direct transmission. Role optimization improves the fixed-photon-number benchmark and closes the ideal-state Bell-violation threshold gap. Nevertheless, the vacuum--$n$-photon protocol retains a lower detection-efficiency threshold for positive key generation while maintaining comparable or higher key rates and communication distances. Hence, our approach provides a practical route toward DI-CKA with current detection technology.

Finally, we discuss the practical realization of the protocol and, as a concrete example, analyze its performance when implemented using spontaneous parametric down-conversion (SPDC) sources. Our results establish the physical encoding of a GHZ state as an important design parameter for all-optical quantum networks and provide a direct method for distributing an encoding tailored to loss-tolerant multipartite Bell tests and near-term device-independent applications. 
%====================================
\section{Detection-Efficiency Thresholds for Parity--CHSH Violations}\label{sec:2}

We investigate the robustness of the photon-number-encoded GHZ-type states $\ket{\Phi_n(q)}$ and $\ket{\Psi_n(q)}$ to detection inefficiency in parity--CHSH tests relevant to DI-CKA. Throughout this work, $n$ is assumed to be even.

The parity--CHSH test was introduced by Ribeiro \emph{et al.} in the context of DI-CKA \cite{ribeiro2018fully} and has subsequently been studied and applied in Refs.~\cite{holz2020genuine,ishihara2025longCKA}. In this test, party A and party B each choose between two dichotomic observables, $A_x$ and $B_y$, with $x,y\in\{0,1\}$, while each of the remaining $n-2$ parties performs a single fixed measurement $C_{1}^{(j)}$. All observables have outcomes in $\{-1,+1\}$. The $n$-party parity--CHSH Bell parameter can be written as \cite{ribeiro2018fully,holz2020genuine}
\begin{align}
S_n
={}&
\langle A_0B_0\rangle
+
\langle A_0B_1\rangle
\nonumber\\
&+
\left\langle
A_1B_0\prod_{j=1}^{n-2}C^{(j)}_1
\right\rangle
-
\left\langle
A_1B_1\prod_{j=1}^{n-2}C^{(j)}_1
\right\rangle ,
\label{eq:parity-chsh-general}
\end{align}
with the classical bound $S_n\leq 2$.

For $n=2$, the product over the $C^{(j)}_1$ observables is understood to be the identity, and Eq.~\eqref{eq:parity-chsh-general} reduces to the standard CHSH expression \cite{clauser1969proposed}.

Our aim is to determine the detection-efficiency thresholds of the two state families without imposing restrictions associated with a particular optical measurement implementation. We therefore allow arbitrary ideal projective qubit measurements. For the parity--CHSH test considered here, the optimizing observables may be chosen in a common $XZ$ plane \cite{ribeiro2018fully,holz2020genuine,ishihara2025longCKA}. Each observable can consequently be parametrized as
\begin{equation}
M(\theta)=\cos\theta\,Z+\sin\theta\,X,
\label{eq:xz-observable-general}
\end{equation}
where $Z$ and $X$ are the standard Pauli operators. The more experimentally realistic displacement-based measurement scheme is considered in Sec.~\ref{sec:long-distanceDI-CKA}.

We model finite detection efficiency by applying an identical and independent amplitude-damping channel $\mathcal A_{\eta}$ of efficiency $\eta$ to each local mode before an otherwise ideal measurement. All measurement rounds, including those affected by photon loss, are retained in the Bell statistics. The Kraus representation of $\mathcal A_{\eta}$ and its action on the relevant Pauli operators are provided in Appendix~\ref{app:exact-parity-chsh-thresholds}.

Because Eq.~\eqref{eq:parity-chsh-general} distinguishes the parties $A$ and $B$ from the remaining parties, the threshold can depend on which physical parties implement these roles. For the permutation-symmetric state $\ket{\Phi_n(q)}$, all assignments are equivalent under the identical-loss model. For $\ket{\Psi_n(q)}$, however, the parties separate into two photon-number classes. The parties assigned the roles $A$ and $B$ may have either equal photon occupations, corresponding to $00$ and $11$ in the two components of the state, or
complementary photon occupations, corresponding to $01$ and $10$. Symmetry within each photon-number class makes all assignments of a given type equivalent. Optimization over the assignment of $A$ and $B$ therefore reduces to comparing these two inequivalent occupation configurations.

For each value of $\eta$, we jointly optimize the state parameter $q$ and all local measurement angles. For $\chi\in\{\phi,\psi\}$ and a given assignment $\lambda$ of the distinguished roles $A$ and $B$, we define the corresponding critical detection efficiency as
\begin{equation}
\eta_{\mathrm{th},\lambda}^\chi
=\inf\left\{\eta\in(0,1):\max_{q,\boldsymbol{\theta}}
S_{n,\lambda}^\chi(\eta,q,\boldsymbol{\theta})>2\right\},
\label{eq:threshold-definition}
\end{equation}
where $\boldsymbol{\theta}$ denotes the collection of local measurement angles. Thus, for the assignment $\lambda$, a parity--CHSH violation is achievable for every $\eta>\eta_{\mathrm{th},\lambda}^\chi$. This optimization is analogous to Eberhard's state-and measurement optimization for two-qubit Bell tests \cite{eberhard1993background}.

\subsection{Complementary-occupation assignment}
\label{subsec:prescribed-role-assignment}

We first consider the assignment in which the parties implementing the two-setting observables $A_x$ and $B_y$ have complementary photon occupations in the two components $\ket{01}^{\otimes n/2}$ and $\ket{10}^{\otimes n/2}$ of $\ket{\Psi_n(q)}$.

For any even $n\geq4$, the joint optimization over the state parameter and all local measurement settings can be solved exactly. The critical detection efficiencies are the unique solutions in $0<\eta<1$ of
\begin{equation}
(\eta_{\mathrm{th}}^\phi)^{n-1}
=2(1-\eta_{\mathrm{th}}^\phi),
\quad (\eta_{\mathrm{th,comp}}^\psi)^{n-1}
=4(1-\eta_{\mathrm{th,comp}}^\psi).
\label{eq:parity-chsh-thresholds}
\end{equation}

This constitutes our first main result. A detailed derivation and proof are provided in Appendix~\ref{app:exact-parity-chsh-thresholds}, together with an explanation of the physical origin of the differing loss tolerances. We show that the difference between the Bell thresholds does not arise from a more favourable preservation of the off-diagonal coherence in the vacuum--$n$-photon state. Instead, it originates from the diagonal correlations entering the parity--CHSH expression under the complementary-occupation assignment.

\begin{figure}[t]
    \centering
    \includegraphics[width=\linewidth]{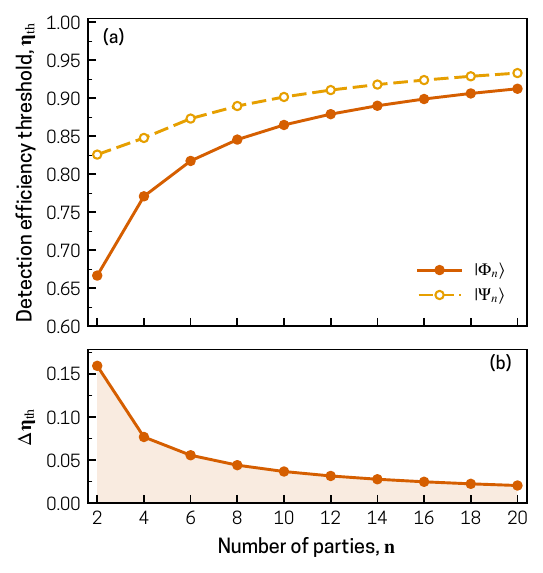}
    \caption{(\textbf{a}) Critical detection efficiencies
    $\eta_{\mathrm{th}}^\phi$ and $\eta_{\mathrm{th,comp}}^\psi$ as functions of the number of parties $n$. For $\ket{\Psi_n(q)}$, the distinguished parties $A$ and $B$ are assigned complementary photon occupations.
    (\textbf{b}) Corresponding reduction in the critical detection efficiency achieved by $\ket{\Phi_n(q)}$, quantified by $\Delta\eta_{\mathrm{th}} =\eta_{\mathrm{th,comp}}^\psi-\eta_{\mathrm{th}}^\phi$.}
    \label{fig:threshold_comparison}
\end{figure}

Each relation in Eq.~\eqref{eq:parity-chsh-thresholds} has a unique solution in $0<\eta<1$. Since the function $f_n(\eta)=\eta^{n-1}/(1-\eta)$ is strictly increasing on this interval and the coefficient for $\ket{\Psi_n(q)}$ is twice that for $\ket{\Phi_n(q)}$, it follows that
$\eta_{\mathrm{th,comp}}^\psi>\eta_{\mathrm{th}}^\phi$ for every even $n\geq4$.
Figure~\ref{fig:threshold_comparison} shows the two critical efficiencies and their difference as functions of $n$. Both thresholds increase with $n$ and approach unity in the large-$n$ limit, while the difference between them gradually narrows. The advantage is therefore largest in the few-party regime. Nevertheless, under the complementary-occupation assignment, $\eta_{\mathrm{th}}^\phi$ remains lower than $\eta_{\mathrm{th,comp}}^\psi$, demonstrating that $\ket{\Phi_n(q)}$ tolerates a greater amount of photon loss in this measurement scenario. While this optimization strategy can be adapted to the bipartite case, the present analysis focuses on the multipartite setting. The $n=2$ thresholds, shown in the figure for completeness, were derived previously in Ref.~\cite{alwehaibi2025tractable}. In that case, each photon-number class of $\ket{\Psi_2(q)}$ contains only one party, so an equal-occupation assignment of the two distinct measurement roles is not available.

\subsection{Equal-occupation assignment}
\label{subsec:optimized-role-assignment}

We now assign the distinguished roles $A$ and $B$ to parties with equal photon occupations in the two components of $\ket{\Psi_n(q)}$. After loss, the two-body diagonal correlation entering the terms $\langle A_0B_0\rangle$ and $\langle A_0B_1\rangle$ is then identical to that obtained for $\ket{\Phi_n(q)}$, up to the equivalent transformation $q\leftrightarrow1-q$ associated with choosing the other photon-number class. By contrast, assigning $A$ and $B$ complementary occupations produces a different diagonal correlation that is independent of $q$. The corresponding correlators are derived explicitly in Appendix~\ref{app:equal-occupation-assignment}.

Choosing $A_1=C_1^{(j)}=X$ and optimizing party B's two observables, gives the exact threshold
\begin{equation}
(\eta_{\mathrm{th,eq}}^\psi)^{n-1}=
2(1-\eta_{\mathrm{th,eq}}^\psi).
\label{eq:psi-optimized-threshold}
\end{equation}
It therefore coincides with the threshold of $\ket{\Phi_n(q)}$ ($\eta_{\mathrm{th,eq}}^\psi=
\eta_{\mathrm{th}}^\phi$). Thus, the separation displayed in Fig.~\ref{fig:threshold_comparison} is a feature of the complementary-occupation assignment and does not represent an intrinsic difference between the role-optimized parity--CHSH thresholds of the two state families. If the distinguished roles can be assigned freely, the detection-efficiency advantage at the level of Bell-inequality violation disappears.

Free assignment of the distinguished roles may, however, be unavailable in an asymmetric network architecture. For example, the two-setting measurements may require active basis selection or fast random-number generation available only at predetermined stations, while the remaining stations implement a single fixed measurement. In such a setting, the assignment of the roles $A$ and $B$ forms part of the physical resources of the protocol, and the complementary-occupation comparison remains operationally relevant.

Moreover, equality of the role-optimized Bell-violation thresholds does not imply equality of the achievable DI-CKA key rates. In addition to the Bell violation used to bound an adversary's information, the conference-key rate depends on the magnitude of the violation, the correlations in the key-generating basis, the reconciliation cost across all parties, and the heralding probability. In particular, assigning $A$ and $B$ equal occupations does not remove the loss-induced key-basis errors between the key-generating party and users in the complementary photon-number class. The assignment that minimizes the Bell-violation threshold therefore need not maximize the complete key-rate expression. The effect of the assignment must be assessed directly at the level of the key rate, as considered in Sec.~\ref{sec:long-distanceDI-CKA}.
\begin{figure*}[t]
    \centering
    \includegraphics[width=\linewidth]{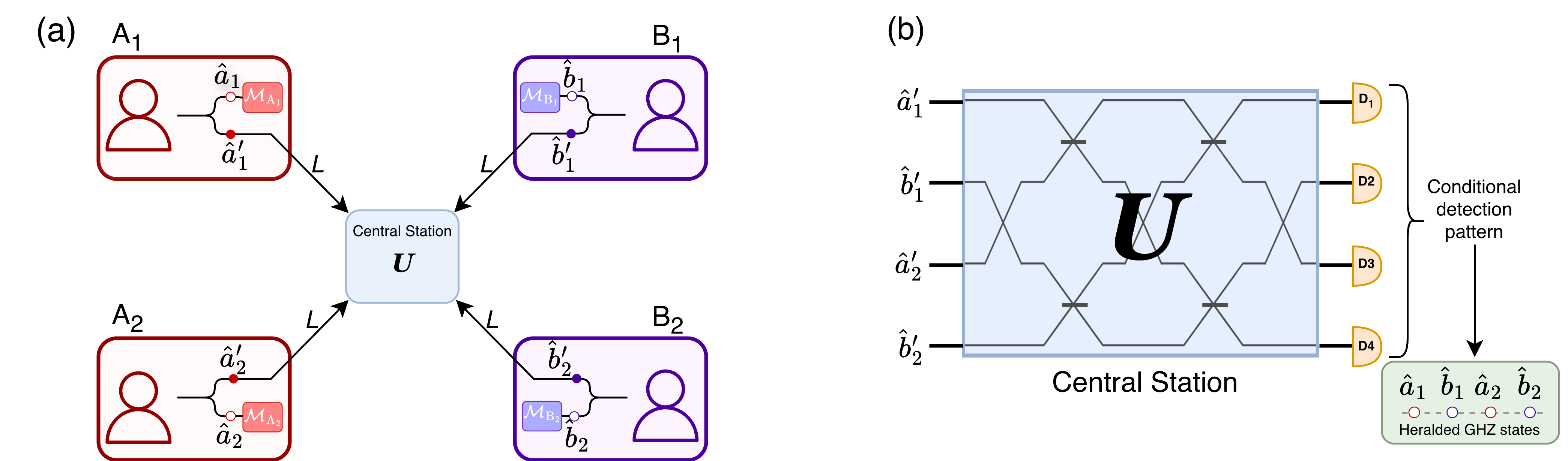}
    \caption{Schematic of the protocol for the heralded distribution of a four-party vacuum--$n$-photon GHZ state. (a) Each party ($\mathsf{A}_1$, $\mathsf{A}_2$, $\mathsf{B}_1$, $\mathsf{B}_2$) holds a local two-mode entangled photon source and transmits one mode ($\hat{a}_1'$, $\hat{b}_1'$, $\hat{a}_2'$, $\hat{b}_2'$) over a channel of length $L$ to a central station, while retaining the other mode ($\hat{a}_1$, $\hat{b}_1$, $\hat{a}_2$, $\hat{b}_2$). (b) The four transmitted modes are combined by a linear-optical unitary $U$ and registered on PNR detectors $D_1$--$D_4$; a conditional detection pattern heralds successful projection of the four retained modes onto the target GHZ state. Conditioned on this heralding signal, a local measurement $\mathcal M$ is subsequently performed on each retained mode.}
    \label{fig:Heralding_scheme}
\end{figure*}
%====================================
\section{Efficient Heralded Distribution of GHZ States}

The preceding section established that $\ket{\Phi_n(q)}$ can offer improved loss tolerance relevant to DI-CKA, with the largest advantage occurring in the few-party regime. The present section introduces a four-user protocol that directly heralds this state while retaining the favourable long-distance scaling $O(\eta_{\text{c}}^{2})$. An extension to arbitrary even ($n>4$), with scaling $O(\eta_{\text{c}}^{n/2})$, is presented in Appendix~\ref{app:extension}.

%==================
\subsection{Protocol Architecture: Ideal Case}

We consider four parties, labelled $\mathsf{A}_1$, $\mathsf{A}_2$, $\mathsf{B}_1$, and $\mathsf{B}_2$, arranged in a star-network topology, as illustrated in Fig.~\ref{fig:Heralding_scheme} (a). Each party locally prepares a two-mode photon-number-path-entangled state across modes $x_i'$ and $x_i$, where $x\in\{a,b\}$ and $i\in\{1,2\}$. $x_i'$ is transmitted to a central station, while $x_i$ is retained locally. The central station applies linear-optical operations followed by photon-number-resolving (PNR) detection. Conditioned on specific detection patterns, a multipartite entangled state is heralded among the modes retained by the users.

The \textit{Type-A} parties, $\mathsf{A}_1$ and $\mathsf{A}_2$, prepare single-photon path-entangled (SPPE) states in the modes $\{a_i,a_i'\}$:
\begin{equation}
\ket{\psi_i}_{a_i a_i'}=\alpha_i\ket{1,0}_{a_i a_i'}
+\beta_i\ket{0,1}_{a_i a_i'},
\label{eq:typeA}
\end{equation}
where the single photon is delocalized between the retained and transmitted modes. The \textit{Type-B} parties, $\mathsf{B}_1$ and $\mathsf{B}_2$, prepare vacuum-pair entangled (VPE) states in the modes $\{b_i,b_i'\}$:
\begin{equation}
\ket{\phi_i}_{b_i b_i'}=\gamma_i\ket{0,0}_{b_i b_i'}
+\delta_i\ket{1,1}_{b_i b_i'},
\label{eq:typeB}
\end{equation}
which describe a superposition of the vacuum and a correlated photon pair. The coefficients satisfy the normalization conditions
$|\alpha_i|^2+|\beta_i|^2=1$ and
$|\gamma_i|^2+|\delta_i|^2=1$.

The global input state is (see Appendix~\ref{app:expansion} for the expanded state form)
\begin{equation}
\ket{\Psi_{\text{in}}}=\ket{\psi_1}_{a_1a_1'}
\otimes\ket{\phi_1}_{b_1b_1'}
\otimes\ket{\psi_2}_{a_2a_2'}
\otimes\ket{\phi_2}_{b_2b_2'}.
\label{eq:global_input}
\end{equation}

At the central station, the transmitted modes are interfered by a network of 50:50 beam splitters. Denoting the output modes by $(d_1,d_2,d_3,d_4)$, monitored by detectors $(D_1,D_2,D_3,D_4)$, respectively, the unitary transformation of the creation operators (see Appendix~\ref{app:interferometer} for full details) is
\begin{equation}
\begin{pmatrix}
d_1^\dagger\\
d_2^\dagger\\
d_3^\dagger\\
d_4^\dagger
\end{pmatrix}
=
U
\begin{pmatrix}
a_1'^\dagger\\
b_1'^\dagger\\
a_2'^\dagger\\
b_2'^\dagger
\end{pmatrix},
\quad
U
=
\frac{1}{2}
\begin{pmatrix}
1 & 1 & 1 & 1\\
1 & -1 & 1 & -1\\
1 & 1 & -1 & -1\\
1 & -1 & -1 & 1
\end{pmatrix}.
\label{eq:app_U_matrix}
\end{equation}

 A successful heralding event requires exactly two photons to arrive at the central station and produce a two-click coincidence, with one photon detected in each of two distinct output modes. Within the bunched-free subspace supported by Eq.~(\ref{eq:global_input}), each such detection pattern projects the transmitted modes onto a coherent superposition of two complementary photon-pair configurations. For example, clicks at detectors $D_1$ and $D_2$ correspond to the output state $\ket{1100}_{d_1d_2d_3d_4}=d_1^\dagger d_2^\dagger\ket{0000}_d$. The corresponding state expressed in terms of the input modes is obtained by back-propagating this detector state through the interferometer unitary. Specifically, 
\begin{align}
\ket{\chi_{12}} &= U^{-1}|1100\rangle = U^{-1} d_1^\dagger d_2^\dagger|0000\rangle_d \nonumber\\
&= \tfrac{1}{2}\big(\ket{1010} - \ket{0101}\big)_{a_1'b_1'a_2'b_2'}+\ket{\tilde\chi},
\label{eq:d1d2_projector}
\end{align}
which defines the measurement projector
$\Pi_{12}=\ket{\chi_{12}}\!\bra{\chi_{12}}$ on the transmitted input modes. Here, $\ket{\tilde\chi}$ is a collection of terms with two photons in one mode.
Projecting the transmitted part of the global input state onto this state yields the following unnormalized state of the retained modes:
\begin{align}
\ket{\Phi_{12}} &= \braket{\chi_{12}}{\Psi_{\text{in}}} \nonumber\\
&= \tfrac{1}{2}\big(\beta_1\gamma_1\beta_2\gamma_2\ket{0000} - \alpha_1\delta_1\alpha_2\delta_2\ket{1111}\big).
\end{align}

Proceeding similarly for all six coincidence patterns yields the set of unnormalized heralded states listed in Table~\ref{tab:2click_heralded_states} (see Appendix~\ref{app:projectors} for details).
Pairs of click patterns yield the same physical state up to a global phase.  

\begin{table}[b]
\centering
\caption{Unnormalized heralded states for each two-click coincidence pattern.}
\label{tab:2click_heralded_states}
\begin{tabular}{c|l}
\toprule
\textbf{Pattern} & \textbf{Heralded state} \\
\midrule
$D_1D_2$ &
$\ket{\Phi_{12}} \propto \beta_1\gamma_1\beta_2\gamma_2\ket{0000} - \alpha_1\delta_1\alpha_2\delta_2\ket{1111}$ \\
$D_3D_4$ &
$\ket{\Phi_{34}}=-\ket{\Phi_{12}}$ \\
\midrule
$D_1D_3$ &
$\ket{\Phi_{13}} \propto \beta_1\delta_1\alpha_2\gamma_2\ket{0110} - \alpha_1\gamma_1\beta_2\delta_2\ket{1001}$ \\
$D_2D_4$ &
$\ket{\Phi_{24}}=-\ket{\Phi_{13}}$ \\
\midrule
$D_1D_4$ &
$\ket{\Phi_{14}} \propto \beta_1\gamma_1\alpha_2\delta_2\ket{0011} - \alpha_1\delta_1\beta_2\gamma_2\ket{1100}$ \\
$D_2D_3$ &
$\ket{\Phi_{23}}=-\ket{\Phi_{14}}$ \\
\bottomrule
\end{tabular}
\end{table}

 This constitutes our second main result. Depending on the observed coincidence pattern, our protocol can herald both fixed-photon-number GHZ-like states, of the type considered in previous work \cite{shimizu2025simple,ishihara2025longCKA}, and the target vacuum-$n$-photon GHZ state. Previous schemes based on identical sources herald only the former class. The heterogeneous source architecture therefore extends the set of directly accessible photon-number encodings to include the computational-basis GHZ state, whose asymmetric photon-number structure gives rise to the enhanced loss tolerance discussed in the previous section. While the present analysis focuses on four parties, the heralded GHZ distribution can be extended to larger even numbers of parties, as described in Appendix~\ref{app:extension}.
%=======================

\subsection{Performance under Photon Loss}
\label{sec:psucc_and_fidelity_under_loss}

We now derive the heralding probabilities of the protocol in the presence of photon loss and analyze the fidelity of the resulting conditional states. Detection patterns heralding the target  vacuum-$n$-photon GHZ state are treated separately from those heralding fixed-photon-number GHZ-like states. For simplicity, the sources are assumed to be symmetric within each type. That is, $\alpha_1=\alpha_2\equiv \alpha$, $\beta_1=\beta_2\equiv \beta$,
$\gamma_1=\gamma_2\equiv \gamma$, $\delta_1=\delta_2\equiv \delta$.

Photon loss in each link is modeled by coupling every transmitted mode $x_i'$ to a vacuum environment mode $E_{x_i}$ through the beam-splitter transformation
\begin{equation}
\ket{1}_{x_i'}\ket{0}_{E_{x_i}}
\longmapsto
\sqrt{\eta_{\text{c}}}\,
\ket{1}_{x_i'}\ket{0}_{E_{x_i}}
+
\sqrt{1-\eta_{\text{c}}}\,
\ket{0}_{x_i'}\ket{1}_{E_{x_i}},
\label{eq:loss_bs}
\end{equation}
followed by tracing over the environment modes. We condition on events in which exactly two distinct detectors each register one photon, while the remaining two detectors register no photons.
%=======================

\begin{figure*}[t]
    \centering
    \includegraphics[width=\linewidth]{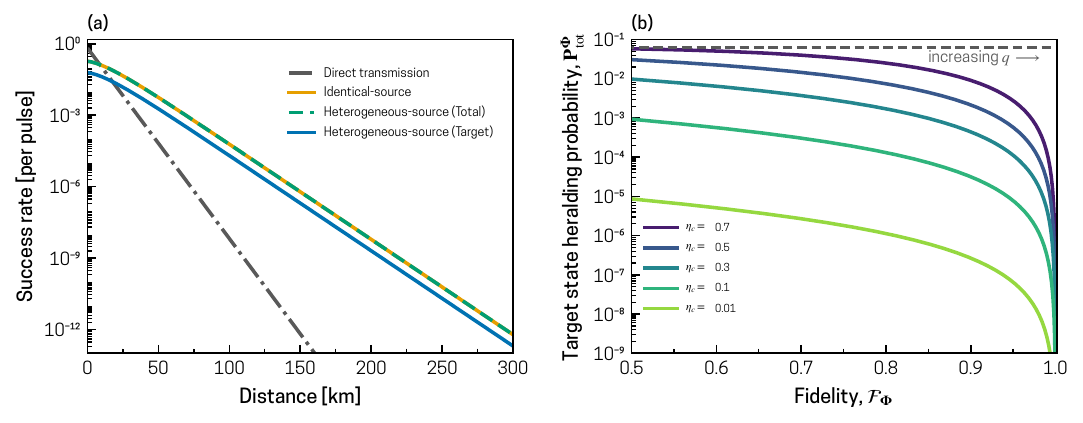}
 \caption{Success rate and fidelity of the target heralded state. (a) Heralding probability per pump pulse as a function of the user-to-central-station link distance, assuming a fiber attenuation of $0.2\,\text{dB/km}$. The orange curve shows the probability of the identical-source protocol, while the dashed green curve, which overlaps it exactly, shows the total probability of our heterogeneous-source protocol. The blue curve shows the contribution associated specifically with heralding the target vacuum-$n$-photon GHZ state. The dashed gray curve denotes the direct-transmission benchmark. The source parameter is $q=0.5$, and the heralding-detector efficiency is $\eta_d=0.9$. (b) Rate--fidelity trade-off for the symmetric condition $|\alpha|^2=|\gamma|^2=q$ at different channel transmittances, corresponding to link distances ranging from approximately $7.7\,\text{km}$ ($\eta_{\text c}=0.7$) to $100\,\text{km}$ ($\eta_{\text c}=0.01$), assuming unity detection efficiency. The horizontal axis shows the fidelity $\mathcal{F}_{\Phi}$, while the vertical axis shows the raw probability of heralding either of the two target detection patterns. Each solid curve is generated parametrically by varying $q$, with the arrow indicating the direction of increasing $q$. The dashed curve connects the unconstrained rate-optimal points $q=q_{\text{opt}}(\eta_{\text c})$.}
\label{fig:Psuccess_fidelity}
\end{figure*}

We first consider the detection pattern $(D_1,D_2)$. The analysis for $(D_3,D_4)$ is identical, and these are the two patterns that herald the target vacuum-$n$-photon GHZ state, as shown in Table~\ref{tab:2click_heralded_states}. The probability of obtaining each individual pattern is
\begin{equation}
\begin{aligned}
P^{\Phi}_{\text{succ}}
=\frac{\eta_{\text{c}}^2}{4}
\Bigl[&|\beta|^4
\bigl(|\gamma|^2+|\delta|^2(1-\eta_{\text{c}})
\bigr)^2\\
&+|\delta|^4\bigl(|\alpha|^2+
|\beta|^2(1-\eta_{\text{c}})\bigr)^2
\Bigr].
\end{aligned}
\label{eq:Psucc_target}
\end{equation}
A detailed derivation, including decomposition into contributions from genuine and loss-induced false-heralding events, is provided in Appendix~\ref{app:psucc_and_fidelity_under_loss_app}. Accepting both $(D_1,D_2)$ and $(D_3,D_4)$ gives the total raw heralding probability associated with the target state,
$P^{\Phi}_{\text{tot}}=2P^{\Phi}_{\text{succ}}$.

The remaining four detection patterns in Table~\ref{tab:2click_heralded_states} are considered next: $(D_1,D_3)$, $(D_2,D_4)$, $(D_1,D_4)$, and $(D_2,D_3)$. These patterns herald fixed-two-photon GHZ-like states. The probability of obtaining each individual pattern is
\begin{equation}
P^{\Psi}_{\text{succ}}
=
\frac{\eta_{\text{c}}^2}{2}
|\beta|^2|\delta|^2
\bigl(
|\alpha|^2+|\beta|^2(1-\eta_{\text{c}})
\bigr)
\bigl(
|\gamma|^2+|\delta|^2(1-\eta_{\text{c}})
\bigr).
\label{eq:P_mixed}
\end{equation}
Since there are four such patterns, their total raw heralding probability is
$P^{\Psi}_{\text{tot}}=4P^{\Psi}_{\text{succ}}$.

For ease of comparison with previous work, we consider the special \textit{symmetric} case
$|\alpha|^2=|\gamma|^2\equiv  q$. The normalization conditions then imply
$|\beta|^2=|\delta|^2=1-q$, and the total target-state heralding probability becomes
\begin{equation}
P^{\Phi}_{\text{tot}}=
\eta_{\text{c}}^2(1-q)^2
\left[1-\eta_{\text{c}}(1-q)
\right]^2.
\label{eq:P_balanced}
\end{equation}
Under the same condition,
$P^{\Psi}_{\text{tot}}=2P^{\Phi}_{\text{tot}}$. Hence, when all six detection patterns are accepted, the total success probability per attempt is $R^{\text{tot}}_{\text{succ}}=3\,P^{\Phi}_{\text{tot}}$. The heralding probability is maximized at $q_{\text{opt}}(\eta_{\text c})=
1-1/2\eta_{\text c}$ (for $\eta_{\text c}>1/2$; see Appendix~\ref{app:psucc_and_fidelity_under_loss_app} for details).

% To compare the resulting heralding probability per pump pulse with the direct-transmission benchmark, we define
% \begin{equation}
% R_{\text{direct}}\equiv \eta_{\text{c}}^n,
% \end{equation}
% and with the identical-source heralding protocol of Ref.~\cite{shimizu2025simple}. 
Defining
\begin{equation}
R_{\text{direct}}\equiv \eta_{\text{c}}^n,
\end{equation}
enables comparison of the resulting heralding probability per pump pulse with the direct-transmission benchmark and the identical-source heralding protocol of Ref.~\cite{shimizu2025simple}. 

For the four-party case considered here, direct transmission therefore scales as $O(\eta_{\text{c}}^4)$, whereas both heralding protocols scale as $O(\eta_{\text{c}}^2)$. Fig.~\ref{fig:Psuccess_fidelity} (a) shows these rates as functions of the user-to-central-station link distance, assuming a standard fiber attenuation. For this comparison, the effective transmission efficiency is $\eta_{\text{c}}=\eta_{\text{ch}}\eta_d$, where $\eta_\text{ch} = 10^{-0.02L}$ is the channel transmittance and $\eta_d$ is the efficiency of the heralding detectors.

Our protocol retains the quadratic rate--loss scaling $O(\eta_{\text{c}}^2)$ characteristic of four-party efficient star-network heralding while enabling the direct distribution of the vacuum-$n$-photon GHZ state. 

At the operating point ($q=1/2$) shown in Fig.~\ref{fig:Psuccess_fidelity} (a), each user prepares a maximally entangled local two-mode state. The probability of heralding the target GHZ state is then slightly lower than in the corresponding identical-source protocol. When all heralded entangled outcomes are included, however, the two protocols have the same total success probability.

In practice, which protocol achieves the higher target-state heralding probability depends on the source technologies and their operating parameters. Nevertheless, this demonstrates that combining different source types preserves the favourable rate--loss scaling while providing direct access to a photon-number encoding that cannot be heralded using an otherwise equivalent identical-source architecture.
%=====================

The conditional state in the presence of photon loss for the detection pattern $(D_1,D_2)$ takes the form (see Appendix~\ref{app:psucc_and_fidelity_under_loss_app} for details)
\begin{equation}
\rho_\phi=\frac{1}{\mathcal N}
\left(p_\Phi\ket{\widetilde{\Phi}}\!\bra{\widetilde{\Phi}}
+\sum_ip_i\ket{\widetilde{\psi}_i}\!\bra{\widetilde{\psi}_i}\right),
\label{eq:rhophi}
\end{equation}
where $\mathcal{N} =P^\Phi_{\text{succ}}$ is the normalization factor. $\ket{\tilde\Phi}=\beta^{2}\gamma^{2}\ket{0000}-\alpha^{2}\delta^{2}\ket{1111}$ is the desired (unnormalized) GHZ component, arising from the coherent superposition of the two genuine loss-free heralding histories, and  the states $\ket{\tilde\psi_i}$ represent loss-induced false-heralding components orthogonal to the desired GHZ subspace.
%====
The two truthful heralding histories do not lose photons to the environment, hence their unnormalized contribution acquires only the common factor $\eta_{\text c}^2$, which does not alter the relative amplitudes within $\ket{\widetilde{\Phi}}$. By contrast, histories involving one or two lost photons leave distinguishable environment records and contribute incoherently after the environment mode is traced out, reducing the fidelity through the admixture of orthogonal false-heralding components.
%====

The quality of the conditional state is quantified using its fidelity with the target four-party GHZ state $\ket{\text{GHZ}_4}=\ket{\Phi_4(1/2)}$
\begin{equation}
\mathcal{F}_{\Phi}
\equiv 
\bra{\text{GHZ}_4}
\rho_\phi
\ket{\text{GHZ}_4}.
\label{eq:fidelity_definition}
\end{equation}
For arbitrary complex source amplitudes, this overlap depends on their relative phase. We assume that this phase, including the detection-pattern-dependent sign, is known and corrected locally. Equivalently, the phase of the target GHZ state may be chosen to maximize the overlap.
The resulting fidelity is
\begin{equation}
\mathcal{F}_{\Phi}=
\frac{\eta_{\text{c}}^2
\left(|\beta|^2|\gamma|^2+|\alpha|^2|\delta|^2
\right)^2}{8\,P^{\Phi}_{\text{succ}}},
\label{eq:fidelity}
\end{equation}
where $P^{\Phi}_{\text{succ}}$ is the single-pattern heralding probability given in Eq.~(\ref{eq:Psucc_target}).

%===========
For the symmetric case (i.e., $|\alpha|^2=|\gamma|^2\equiv  q$, with
$|\beta|^2=|\delta|^2=1-q$), the fidelity reduces to
\begin{equation}
\mathcal{F}_{\Phi}
=\left[\frac{q}
{1-(1-q)\eta_{\text{c}}}
\right]^2.
\label{eq:F_balanced}
\end{equation}

Together with Eq.~(\ref{eq:P_balanced}), this expression reveals the trade-off between heralding probability and fidelity shown in Fig.~\ref{fig:Psuccess_fidelity} (b). For a lossy channel, the fidelity increases monotonically with $q$ and approaches unity as $q\to1$. In the lossless limit, $\eta_{\text c}=1$, the fidelity is unity for every $q>0$. Pushing $q$ towards the high-fidelity regime, however, reduces the raw heralding probability. 

Genuine multipartite entanglement (GME) is certified by the projector-based GHZ witness whenever $\mathcal{F}_{\Phi}>1/2$, where $1/2$ is the maximum fidelity between a multipartite GHZ state and any biseparable state~\cite{guhne2007toolbox,guhne2009entanglement}. It is worth noting that the GME witness here is state-based and is distinct from the parity--CHSH certification of nonlocal correlations. At fixed $q$, the fidelity decreases monotonically with distance and approaches $\mathcal{F}_{\Phi}\approx q^2$. Thus, choosing $q>1/\sqrt{2}$ ensures that the heralded state remains genuinely multipartite entangled at arbitrarily long distances. This robustness arises because every accepted event requires the detection of two transmitted photons, so all terms in the unnormalized heralded state contain a common factor $\eta_{\text c}^{2}$. This factor cancels upon normalization, while the remaining loss-dependent coefficients approach finite values as $\eta_{\text c}\to0$ (see Appendix~\ref{app:psucc_and_fidelity_under_loss_app} for details). Consequently, the conditional fidelity approaches the limit $q^2$ for $q>0$, even though the heralding probability continues to decrease as $O(\eta_{\text c}^{2})$. The range of the protocol is thus limited not by fidelity but by the rate falling below the dark-count floor. The fidelity, $\mathcal{F}_{\Psi}$, of heralding the other states is discussed in Appendix~\ref{app:psucc_and_fidelity_under_loss_app}.   
%============

%====================================
\section{Long-distance DI-CKA}\label{sec:long-distanceDI-CKA}

In this section, we use the heralded GHZ state distribution scheme developed earlier to construct and analyze a long-distance DI-CKA protocol. We consider four legitimate parties, $A$, $B$, $C^{(1)}$, and $C^{(2)}$ who seek to establish a common secret key. Here, $A,B,C^{(1)},C^{(2)}$ denote measurement roles, whereas
$\mathsf{A}_1,\mathsf{A}_2,\mathsf{B}_1,\mathsf{B}_2$ label the physical parties according to their source types. The assignment of physical parties to measurement roles may vary between the configurations considered below. Each party is located $L$ kilometers from the central station. Two parties use Type-A sources and two use Type-B sources, where the assignment of these physical parties to the measurement roles is specified in the comparisons below. The GHZ states are distributed among the parties in the way described in the previous section. Then, each party performs measurements on the distributed GHZ states. The parties repeat the above procedure for a sufficient number of rounds. Let $A_x$ denote the measurement performed by $A$ with a binary input $x \in \{ 0, 1\}$ and a binary output $a \in \{ 0, 1 \}$. Similarly, $B_{y}$ denotes $B$'s measurements with an input $y \in \{0, 1, 2\}$ and an output $b \in \{0,1\}$. Let $C_{z_j}^{(j)}$ denote $C^{(j)}$'s measurements with a binary input $z_j \in \{ 1, 2 \}$ and a binary output $c_j \in \{ 0, 1\}$ for $j = 1, 2$. Rounds where the inputs $x = 0$ and $y = z_1 = z_2 = 2$ are chosen are classified as key-generation rounds, and rounds where the party $A$ ($B$) chooses $x \, (y) \in \{0, 1  \}$ and the remaining parties choose $z_1 = z_2 = 1$ are classified as Bell-test rounds and used to test for a violation of the parity--CHSH inequality. Furthermore, for the key-generation rounds, one of the legitimate parties, e.g., the party $A$, performs noisy preprocessing~\cite{Ho2020}. In noisy preprocessing, the party flips each raw key bit with probability $p_\text{noisy}$. This procedure reduces the amount of information on $A$'s system which an eavesdropper (Eve) possesses, and enhances the robustness of the DI-CKA protocol against imperfections. Using a subset of the key-generation rounds and the Bell-test rounds, the parties calculate a DI-CKA key rate. If the calculated key rate is positive, they perform error correction and privacy amplification, establishing a shared secret key.

The asymptotic DI-CKA key rate $r$ can be expressed as~\cite{ribeiro2018fully}
\begin{equation}\label{eq:keyrate}
\begin{split}
    r &= P_\text{herald}[H(A|E) \\
    &- \max \{ H(A|B), H(A|C^{(1)}), H(A|C^{(2)}) \}],
\end{split}
\end{equation}
where $H(A|E)$ represents a conditional von Neumann entropy of $A$'s system given Eve's information, the second term corresponds to the maximal error correction cost for $A$ and the other parties, and $P_\text{herald}$ represents the probability that the GHZ states are successfully distributed among the legitimate parties. The second term in Eq.~(\ref{eq:keyrate}) can be calculated from a probability distribution of measurement results for the key-generation rounds. On the other hand, the first term is estimated from a violation of the parity--CHSH inequality obtained for the Bell-test rounds. A tight analytical lower bound on $H(A|E)$ for the parity--CHSH inequality with noisy preprocessing exists and is given by \cite{Ishihara2026Enhancing}

\begin{equation}
\begin{split}
    H(A|E) &\geq 1-h\left(\frac{1+\sqrt{S_4^2/4-1}}{2}\right) \\&+ h\left(\frac{1 + \sqrt{1-(1-p_\text{noisy})p_\text{noisy}(8-S_4^2)}}{2}\right),
\end{split}
\end{equation}
where $S_4$ is the observed value of the parity--CHSH parameter and $h$ denotes the binary Shannon entropy. In this DI-CKA protocol, the legitimate parties share a secret key from events where the detection pattern $(D_1, D_2)$ occurs. Therefore, the heralding probability $P_\text{herald}$ is equivalent to the probability that the detection pattern $(D_1, D_2)$ occurs, i.e., $P_\text{herald} = P^\Phi_{\text{succ}}$ where $P^\Phi_{\text{succ}}$ is given in Eq.~(\ref{eq:Psucc_target}) in the absence of dark counts.
\begin{figure*}[t]
    \centering
    \includegraphics[width=\linewidth]{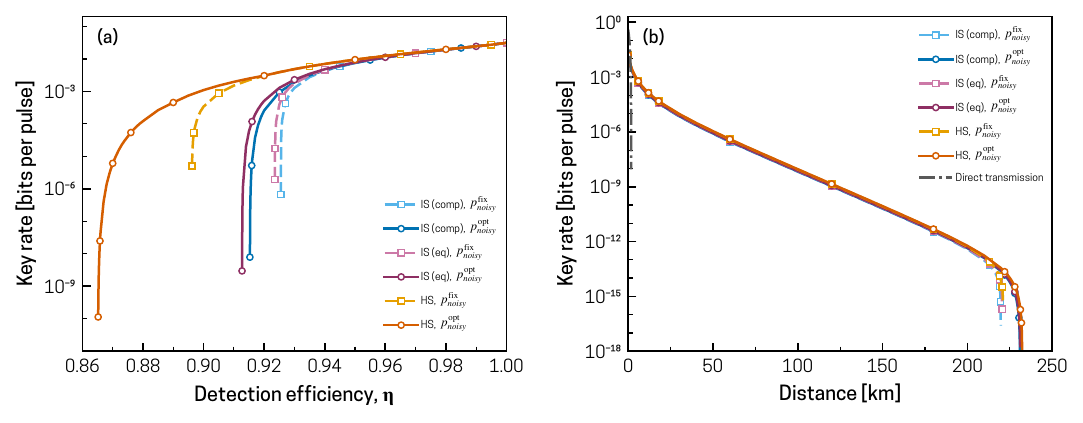}
 \caption{Key rates with ideal Pauli measurements. (a) Key rate versus detection efficiency $\eta$. The orange curves represent the heterogeneous-source (HS) protocol. The blue and purple curves represent the identical-source (IS) protocol with complementary-occupation (comp) and equal-occupation (eq) assignments of the two-setting roles, respectively. Square (circular) markers indicate that $p_\text{noisy}=0$ ($p_\text{noisy}$ is optimized), labeled $p_\text{noisy}^\text{fix}$ ($p_\text{noisy}^\text{opt}$). We numerically optimize the measurement angles and the relevant source parameters. We use $\eta_c = 1$ and $p_\text{DC} = 10^{-8}$. (b) Key rate versus distance $L$ for the same protocols and assignments. The gray dashed line represents the direct-transmission protocol. We use $\eta = 0.97$ and $p_\text{DC} = 10^{-8}$.}
\label{fig:Comparison_with_ideal_measurements}
\end{figure*}
In the following, we calculate the DI-CKA key rate using two models for the parties' measurements: ideal Pauli measurements in the $XZ$ plane and displacement-based measurements. Since distributed quantum states in this DI-CKA protocol are encoded in the Fock basis, the ideal Pauli measurements are difficult to implement in an experiment. On the other hand, the displacement-based measurement is realized with an optical displacement operation followed by an on-off detector, which can be implemented in practice with current technology. Note that we can implement the Pauli measurements in the Fock basis by using matter-based systems~\cite{Ashhab2007} or Dolinar-inspired detectors~\cite{Takeoka2005, Takeoka2006}, although these approaches are more challenging to implement than the displacement-based measurements.

We consider the following experimental imperfections. We account for effective channel loss $\eta_c$ as defined before;
the detection efficiency of the legitimate parties, $\eta$; the dark-count probability of the detectors at the central station, $p_\text{DC}$. For simplicity, we assume that $\alpha_1 = \alpha_2 = \alpha, \beta_1 = \beta_2 = \beta, \gamma_1 = \gamma_2 = \gamma$, and $\delta_1 = \delta_2 = \delta$ in this section. In Appendix~\ref{app:keyratesimulation}, we describe the simulation model used to calculate the key rate.

\subsection{Ideal Pauli Measurement}
We first consider the ideal Pauli measurements in the $XZ$ plane. This measurement can be expressed as Eq.~(\ref{eq:xz-observable-general}). For the Bell-test rounds, the party $A$ ($B$) performs $A_0 = M(0)$ and $A_1 = M(\theta_1)$ ($B_0 = M(\theta_2)$ and $B_1 = M(\theta_3)$), and the remaining parties perform $ C_1^{(j)} = M(\theta_4)$ for $j \in \{1,2 \}$. For the key-generation rounds, all the parties perform the $Z$ measurement, i.e., $A_0 = B_2 = C_2^{(1)} = C_2^{(2)} = M(0)$. We optimize the measurement angles $\theta_1, \ldots, \theta_4$ to maximize the key rate.

We first compare the local detection efficiencies required to obtain a positive key rate. Fig.~\ref{fig:Comparison_with_ideal_measurements} (a) shows the results for our heterogeneous-source protocol in orange and for the identical-source protocol of Ref.~\cite{ishihara2025longCKA} in blue and purple. In the latter protocol, every party is equipped with a Type-A source, and the distributed state is the fixed-photon-number GHZ state $\ket{\Psi_4}$. Since the modes of this state differ in photon occupation within each component, its parity--CHSH performance depends on which parties are assigned the two-setting roles, as shown in Sec.~\ref{sec:2}. We therefore consider two assignments: the complementary-occupation assignment of Ref.~\cite{ishihara2025longCKA}, labeled IS (comp), in which $A$ and $B$ have complementary photon occupations in each component, and the equal-occupation assignment, labeled IS (eq), in which $A$ and $B$ have equal photon occupations in each component. Square markers denote results without noisy preprocessing, $p_{\mathrm{noisy}}=0$ (labeled $p_\text{noisy}^\text{fix}$), whereas circular markers denote results obtained by optimizing $p_{\mathrm{noisy}}$ to maximize the key rate (labeled $p_\text{noisy}^\text{opt}$). The source parameters $\alpha$, $\beta$, $\gamma$, and $\delta$ are also optimized numerically. Without noisy preprocessing, the detection-efficiency thresholds of IS (comp) and IS (eq) are approximately $92.55\%$ and $92.35\%$, respectively, whereas the heterogeneous-source architecture lowers the threshold to $89.61\%$. When noisy preprocessing is optimized, the corresponding thresholds decrease to $91.53\%$, $91.27\%$, and $86.51\%$, respectively. Relative to the identical-source protocol with the equal-occupation assignment, the heterogeneous-source protocol therefore lowers the threshold by $2.7$ and $4.8$ percentage points without and with noisy preprocessing, respectively. This reduction applies at every user and therefore provides a substantial efficiency margin in a multipartite implementation.
\begin{figure*}[t]
    \centering
    \includegraphics[width=\linewidth]{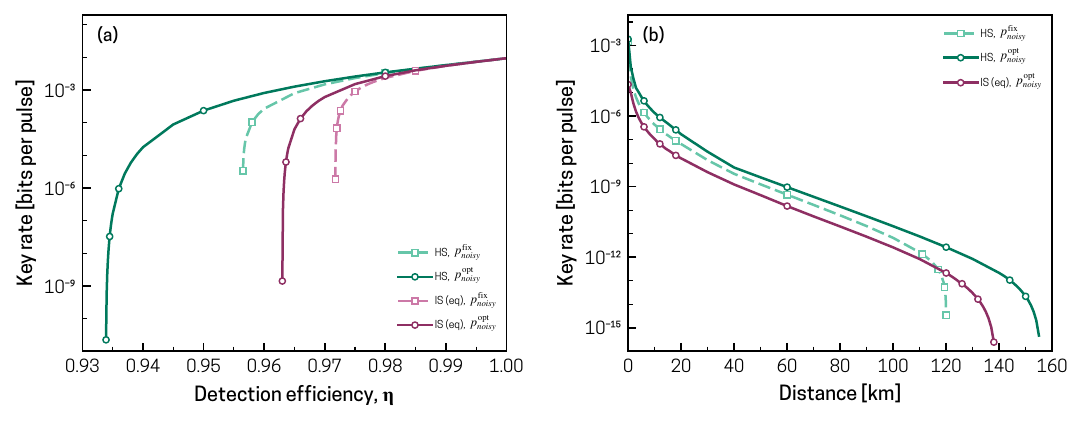}
\caption{Key rates with displacement-based measurements. (a) Key rate versus detection efficiency $\eta$ for the heterogeneous-source (HS) protocol in green and the identical-source (IS) protocol with equal-occupation (eq) assignment in purple. Square (circular) markers indicate that $p_\text{noisy}=0$ ($p_\text{noisy}$ is optimized), labeled $p_\text{noisy}^\text{fix}$ ($p_\text{noisy}^\text{opt}$). We numerically optimize the displacement amplitudes and the relevant source parameters. We use $\eta_c = 1$ and $p_\text{DC} = 10^{-8}$. (b) Key rate versus distance $L$ for the same protocols and assignments. We use $\eta = 0.97$ and $p_\text{DC} = 10^{-8}$.}
\label{fig:HS_keyrate_with_displacement}
\end{figure*}
This remaining advantage is not determined by the parity--CHSH violation threshold alone. Under equal-occupation assignment, the two encodings have the same ideal-Pauli Bell-violation threshold (Sec.~\ref{sec:2}), but Eq.~(\ref{eq:keyrate}) also depends on the largest error-correction cost between $A$ and the other parties. In $\ket{\Psi_4}$, the parties divide into two photon-occupation classes. Regardless of which class contains $A$, two of the other three parties belong to the opposite class, and exactly one member of each such pair carries a photon in each component. For the ideal two-component state under independent loss, their key bits disagree with probability $1-\eta$ after the expected bit flip, independently of the component weights. In $\ket{\Phi_4}$, by contrast, all parties share the vacuum component, which is unaffected by loss; loss-induced key errors arise only from the four-photon component and can be reduced by adjusting its weight. The full loss-density-matrix comparison in Appendix~\ref{app:full-loss-density-matrix-comparison} makes this difference in loss response explicit. This helps explain why equal Bell-violation thresholds do not lead to equal key-generation thresholds.

In Fig.~\ref{fig:Comparison_with_ideal_measurements} (b), we plot key rates against channel distance $L$ between each party and the central station. Again, the orange squares and circles represent key rates of the heterogeneous-source DI-CKA protocol without and with noisy preprocessing, respectively. The corresponding blue and purple curves represent the identical-source protocol with complementary- and equal-occupation assignments. For comparison, we plot key rates of a direct-transmission protocol, where the vacuum-$n$-photon GHZ state is generated at the central station and distributed among the parties, as the gray dashed line. While the direct-transmission protocol supports a range of only a few kilometers, both heralded protocols can achieve communication distances of over 200 km. Noisy preprocessing further increases their maximal communication distances. The equal-occupation assignment slightly improves the key rates of the identical-source protocol, while the heterogeneous-source protocol retains a slight advantage. These rates are close because Fig.~\ref{fig:Comparison_with_ideal_measurements} (b) assumes a high detection efficiency of $\eta=97\%$. The threshold comparison in Fig.~\ref{fig:Comparison_with_ideal_measurements} (a) shows a more pronounced distinction. Only the heterogeneous-source protocol produces a positive key rate for detection efficiencies between approximately $86.51\%$ and $91.27\%$. Thus, our protocol markedly lowers the required detection efficiency while maintaining key rates comparable to those of the identical-source protocol at high detection efficiency.

\subsection{Displacement Measurement}
\label{subsec:displacement}
Next, we consider the displacement-based measurement. The measurement operators of this measurement are $\{\ketbra{\xi}{\xi} , I - \ketbra{\xi}{\xi} \}$,
where $\ket{\xi}$ is a coherent state with a complex amplitude $\xi$ and $I$ is the identity operator. We model the measurements of the legitimate parties as follows
\begin{align}
    M_\text{dis}(\xi) &= 2\ketbra{\xi}{\xi} - I.
\end{align}
The party $A$ ($B$) performs $A_0 = M_\text{dis}(0)$ and $A_1 = M_\text{dis}(\xi_1)$ ($B_0 = M_\text{dis}(\xi_2)$ and $B_1 = M_\text{dis}(\xi_3)$), and the remaining parties perform $C_1^{(1)} = C_1^{(2)} = M_\text{dis}(\xi_4)$ for the Bell-test rounds. All the parties perform $M_\text{dis}(0)$ for the key-generation rounds. We optimize the amplitudes $\xi_1, \ldots, \xi_4$ to maximize the key rate.

In Fig.~\ref{fig:HS_keyrate_with_displacement}, we plot key rates for the displacement-based measurement. Fig.~\ref{fig:HS_keyrate_with_displacement} (a) shows key rates against the detection efficiency $\eta$. The green curves represent our heterogeneous-source protocol, while the purple curves represent the identical-source protocol with equal-occupation assignment. Square (circular) markers represent key rates without (with) noisy preprocessing. The relevant source parameters are optimized numerically for each protocol. In Fig.~\ref{fig:HS_keyrate_with_displacement} (b), we plot key rates against channel distance $L$. The threshold detection efficiencies are higher and the maximal communication distances shorter than those for ideal Pauli measurements in Fig.~\ref{fig:Comparison_with_ideal_measurements}. This is because the achievable parity--CHSH value is smaller with the displacement-based measurement than with the ideal Pauli measurement. Noisy preprocessing significantly enhances the robustness of both protocols against experimental imperfections. Moreover, with the displacement-based measurements, our protocol achieves a positive key rate with about $93.4\%$ detection efficiency and communication distance of over 150 km, substantially exceeding that of the direct-transmission protocol with the ideal Pauli measurements. Under channel loss, the heralded state need not retain the permutation symmetry of the ideal $\ket{\Phi_4}$ state, and optimizing the measurement-role assignment yields a marginal further improvement in key rate and communication distance. The identical-source protocol with the equal-occupation assignment requires detection efficiencies of approximately $97.2\%$ and $96.3\%$ without and with noisy preprocessing, respectively, and reaches a maximal communication distance of approximately 138 km with noisy preprocessing.

The equal-occupation assignment also changes the security analysis available for the identical-source protocol. With displacement-based measurements and this assignment, it exhibits a parity--CHSH violation and a positive key rate that can be certified using the analytical entropy bound above. Thus, in this setting, it no longer requires the direct numerical optimization of the conditional von Neumann entropy used in previous work~\cite{ishihara2025longCKA}. Taken together, these results show that our heterogeneous-source protocol achieves a lower detection-efficiency threshold, higher key rates at imperfect detection efficiencies, and a longer communication distance, while the equal-occupation assignment also improves the existing identical-source protocol. Both protocols now connect an experimentally accessible Bell violation to an analytical key-rate bound.

%=================================================
\section{Physical Implementation}
\label{sec:physical_implementation}

The Type-A SPPE state used in both the identical-source and heterogeneous-source protocols is comparatively straightforward to prepare. A single photon is injected into a variable beam splitter, whose transmittance and relative phase determine the state coefficients. Suitable single photons may be obtained directly from on-demand quantum emitters \cite{somaschi2016near,senellart2017high} or heralded using SPDC sources \cite{mosley2008heralded,donati2014,guo2017parametric}. Further discussion of the realization of the identical-source protocol is given in Ref.~\cite{shimizu2025simple}.

Preparing the Type-B VPE state for the heterogeneous-source protocol is more demanding. The most direct realization is a weak two-mode squeezed-vacuum (TMSV) state generated by spontaneous parametric down-conversion (SPDC). Its vacuum and single-pair components have the required form, but the source also contains unwanted multipair contributions. Although reducing the SPDC gain suppresses their absolute probabilities, it does not necessarily suppress them relative to the desired contribution in the present multiuser protocol. In particular, the desired emission of one pair from each of the two Type-B sources and the unwanted emission of two pairs from a single Type-B source occur at the same order in the SPDC gain. Multipair contamination can therefore produce false-heralding events and significantly reduce the conditional-state fidelity.

Appendix~\ref{app:performance_with_spdc} presents a detailed analysis of the SPDC implementation, deriving the heralded state and its fidelity and evaluating its parity--CHSH violation with realistic measurements. In the lossless heralding limit, the state contains the desired coherence between $\ket{0000}$ and $\ket{1111}$, together with $\ket{1210}$ and $\ket{1012}$ leakage terms produced when one Type-B source emits two pairs. For identical Type-B sources, the maximum lossless fidelity with $\ket{\Phi_4(q)}$ is $\mathcal F_{\text{SPDC}}^{\max}(q)=(2-q)/2$: it is $3/4$ for a maximally entangled target and approaches unity as the target becomes vacuum dominated. Despite the multipair leakage, the heralded state can violate parity--CHSH with displacement-based measurements. The leakage distinguishes the Type-A and Type-B modes, making the measurement-role assignment consequential. Assigning both two-setting roles to Type-A parties gives a detection-efficiency threshold of $0.8932$ with identical Type-B gains. Assigning one role to a Type-A party and the other to a Type-B party gives a higher threshold of $0.9487$, even when unequal Type-B gains are allowed.

One way to suppress these unwanted Fock sectors is to truncate the TMSV state using quantum-scissors operations or cascaded SPDC sources~\cite{pegg1998optical,xue2025photon}. Alternatively, the state can be prepared in an event-ready manner using the efficient heralding scheme introduced in Ref.~\cite{alwehaibi2025tractable}. These approaches incur additional resource overhead in terms of ancillary states and conditional detection channel for each Type-B user.

A more technologically demanding route is provided by coherently driven quantum emitters. Sequential resonant excitation of a two-level emitter has been experimentally used to generate photon-number Bell states of the form $c_0\ket{0,0}+c_1\ket{1,1}$ across two temporal modes~\cite{wein2022photon,vajner2025exploring}. For use in the present protocol, these temporal modes can be coherently routed into separate spatial paths using a synchronized electro-optic switch or another active time-to-space demultiplexer. Ideally, this mode conversion preserves the photon-number entanglement and introduces only a controllable relative phase. Such emitter-based approaches avoid the intrinsic geometric multipair tail of TMSV sources and provide a potential route towards on-demand VPE generation, albeit with greater experimental complexity.

%==========================

\section{Conclusion and Outlook}
\label{sec:conclusion}

We have shown that the loss tolerance of photonic GHZ states is determined not only by their abstract entanglement structure but also by their photon-number encoding and how measurement roles are assigned to the physical modes. For arbitrary even $n$, we derived exact analytical conditions for the critical detection efficiencies of vacuum--$n$-photon and fixed-photon-number GHZ states in parity--CHSH tests. Under the conventional role assignment, the vacuum--$n$-photon encoding tolerates more loss for every $n$, most markedly in the few-party regime. For ideal Pauli measurements, however, assigning the two-setting roles to parties with equal photon occupations closes this Bell-threshold gap. The two encodings nevertheless differ in key generation, which involves the outcomes of all parties. The vacuum--$n$-photon encoding confines loss-induced key-bit errors to the $n$-photon component, allowing them to be suppressed in the vacuum-biased regime relevant near threshold, a feature absent for the other encoding.

Building on this, we introduced two routes to improving the loss tolerance of DI-CKA with heralded photonic GHZ states: a heterogeneous-source scheme that directly heralds the vacuum--$n$-photon encoding, and an optimized measurement-role assignment that improves existing fixed-photon-number schemes. The former overcomes the restriction of an otherwise equivalent identical-source architecture while retaining the favourable scaling $O(\eta_{\text c}^{n/2})$. The same central measurement also heralds the fixed-photon-number states of previous work, and, within our loss model, genuine multipartite entanglement persists at any finite distance, with the practical reach set by the heralding rate and the dark-count floor rather than by unavoidable fidelity degradation.

For four users, the protocol achieves positive DI-CKA key rates at a detection efficiency of $86.51\%$ with ideal Pauli measurements, compared with $91.27\%$ for the role-optimized identical-source protocol, and at about $93.4\%$ with experimentally accessible displacement-based measurements, over distances exceeding $150~\mathrm{km}$. With displacement-based measurements, role optimization also enables parity--CHSH-based certification for the identical-source protocol, allowing an analytical Bell-based entropy bound in place of direct numerical entropy optimization, although our protocol retains a lower detection-efficiency threshold and a longer communication distance. Finally, an SPDC-based implementation of the heterogeneous-source protocol was analyzed. In our low-gain model, multipair emission imposes a gain-independent fidelity ceiling for a maximally entangled state. Nevertheless, the heralded state can still violate the parity--CHSH inequality with displacement-based measurements, and its source-induced asymmetry makes the role assignment consequential. Together, these results establish measurement-role assignment, alongside photon-number encoding and source architecture, as a key design parameter whenever asymmetries in the encoding, sources, channels, or detection efficiencies break the permutation symmetry of the network.

Our results suggest several directions for future work in all-photonic multipartite systems and quantum networks. An immediate extension is to determine whether the key-generation advantage of the vacuum--$n$-photon encoding and the role dependence identified here persist for other multipartite Bell inequalities, and whether alternative inequalities provide more favourable scaling with the number of parties \cite{mermin1990extreme,ardehali1992bell,svetlichny1987distinguishing}. Beyond DI-CKA, the heralded states considered here may provide useful resources for distributed quantum sensing, global phase and privacy-preserving sensing \cite{ueda2025quantum,zang2026quantum,proctor2018multiparameter,shettell2022cryptographic,de2025anonymous}, and distributed quantum computation, where multipartite entanglement carried over lossy photonic links mediates stabilizer measurements between separated processor cells \cite{nickerson2013topological,nickerson2014freely,de2024thresholds, shimizu2025simple}. A further direction is to explore integrated photonic implementations of our protocol \cite{wang2020integrated,pont2024high}, enabling on-chip heralding of single-rail GHZ states with potential applications in compact multiparty quantum communication and photonic quantum information processing \cite{llewellyn2020chip,pont2024high,adcock2019programmable}. Finally, the central station admits a richer set of detection outcomes than the two-click coincidence patterns analyzed here. Because the total input state has even photon-number parity, conditioning in the ideal loss-free setting on exactly one detected photon or on two photons bunched in the same output mode projects the retained modes onto odd- or even-parity sectors, respectively. Exploring these and other detection outcomes, together with alternative combinations of source types, may therefore provide access to a broader class of multipartite entangled states and heralded quantum-network primitives.  

\begin{acknowledgments}
This work was supported by UK Research and Innovation Future Leaders Fellowship (project MR/W011794/1) and the National Company of Telecommunications and Information Security (NTIS). M.I., W.R., and M.T. acknowledge support from JST SPRING, Grant No. JPMJSP2123, JST Moonshot R\&D, Grant No. JPMJMS256E, SQAI (COI-NEXT): JPMJPF2221, JST CRONOS, Grant No. JPMJCS24N6, and JST ASPIRE, Grant No. JPMJAP2427. Claude (Opus 4.6) and ChatGPT (GPT-5.6 Sol) assisted with the measurement optimization presented in Appendix~\ref{app:measurement_optimization}, which supported the derivation of Eq.~(\ref{eq:parity-chsh-thresholds}), and helped identify the importance of measurement-role assignment. These tools also assisted with plotting code and manuscript editing. All LLM-assisted content was independently reviewed and verified by the authors, who take full responsibility for the paper’s content.
\end{acknowledgments}

%======================
\appendix
\renewcommand{\appendixname}{APPENDIX}
\crefalias{section}{appendix}

%===================================

\section{Exact Parity--CHSH Loss Thresholds}
\label{app:exact-parity-chsh-thresholds}

Here, we investigate the performance of two single-rail states under loss and finite detection efficiency in the context of the parity--CHSH test used in device-independent conference-key agreement (DI-CKA) \cite{ribeiro2018fully, holz2020genuine, murta2020quantum}.

We compare the vacuum--$n$-photon entangled state
(Eq.~\eqref{eq:phi_n}) and the fixed-photon-number state
(Eq.~\eqref{eq:psi_n}) and derive their exact detection-efficiency thresholds for the parity--CHSH test. For $\ket{\Psi_n(q)}$, we consider separately the two inequivalent assignments in which the distinguished parties $A$ and $B$ have either equal or complementary
photon occupations. The derivation is valid for every even number of parties $n\geq4$ and includes a global proof that no choice of local projective measurements can produce a violation below the stated thresholds.

For a state family $\chi$ and an assignment $\lambda$ of the distinguished roles, we define
\begin{equation}
\eta_{\mathrm{th},\lambda}^\chi
=\inf\left\{\eta\in(0,1):
\max_{q,\boldsymbol{\theta}}
S_{n,\lambda}^\chi(\eta,q,\boldsymbol{\theta})>2
\right\},
\label{eq:app_threshold-definition}
\end{equation}
where $\boldsymbol{\theta}$ denotes the collection of local measurement angles. Thus, $q$ is not fixed in advance: for every efficiency $\eta$, both the state weight $q$ and all measurement angles are optimized. For the permutation-symmetric state $\ket{\Phi_n(q)}$, all assignments are equivalent, and we suppress the assignment label by writing its threshold as $\eta_{\mathrm{th}}^\phi$. For $\ket{\Psi_n(q)}$, we use
$\lambda\in\{\mathrm{eq},\mathrm{comp}\}$ to denote the equal- and complementary-occupation assignments, respectively.

We recall that the inequality is given by Eq.~(\ref{eq:parity-chsh-general}). The relevant projective observables for DI-CKA may be chosen in a common $XZ$ plane \cite{ribeiro2018fully, holz2020genuine, murta2020quantum}. We therefore write the measurement operator as
\begin{equation}
M(\theta)=\cos\theta\, Z+\sin\theta\, X.
\label{eq:xz-observable-general_app}
\end{equation}
For compactness, we define
\begin{equation}
u_M\equiv \cos\theta_M,
\qquad
v_M\equiv \sin\theta_M.
\label{eq:uv-angle-definition}
\end{equation}

\subsection{States and loss model}

We label the parties by
\[A,\quad B,\quad C^{(1)},\ldots,C^{(n-2)},\]
where party A and party B each have two measurement settings, while the remaining $n-2$ parties each use one fixed setting in the parity--CHSH test. 

Detection inefficiency is modelled by an independent amplitude-damping channel
$\mathcal A_\eta$ of efficiency \(\eta\) on every mode. Its Kraus
operators are
\begin{equation}
K_0=\ket{0}\!\bra{0}+\sqrt{\eta}\,\ket{1}\!\bra{1},
\qquad K_1=\sqrt{1-\eta}\,\ket{0}\!\bra{1}.
\label{eq:loss-kraus-threshold}
\end{equation}
In the Heisenberg picture, the dual channel acts on the relevant Pauli operators as
\begin{equation}
\mathcal{A}_\eta^\dagger(X)=\sqrt{\eta}\,X,
\qquad
\mathcal{A}_\eta^\dagger(Z)=\eta Z+(1-\eta)I.
\label{eq:heisenberg-loss}
\end{equation}

\subsection{Correlators after loss}
The two branches of either state are bitwise complementary, differing
in every mode. It follows that
a Pauli string containing $X$ on only a subset of the modes has zero expectation value. A diagonal contribution requires only
$Z$'s and identities, whereas an off-diagonal contribution between
the two branches requires flipping all $n$ modes. Hence, the only quantities entering the parity--CHSH expression are a two-body
$ZZ$ correlator, a full $Z^{\otimes n}$ correlator, and the full
$X^{\otimes n}$ coherence.

For either family, the full coherence is
\begin{align}
G_n^X(\eta,q)&\equiv 
\left\langle X^{\otimes n}\right\rangle_{\eta}
\nonumber\\
&=\left(\sqrt{\eta}\right)^n
\left\langle X^{\otimes n}\right\rangle_{\text{in}}
\nonumber\\
&=2\eta^{n/2}\sqrt{q(1-q)}.
\label{eq:gn_X}
\end{align}

The two-body $ZZ$ and full $Z^{\otimes n}$ correlators are
state-dependent. For now, we write them as
\begin{equation}
G_2\equiv \langle Z_AZ_B\rangle,
\qquad
G_n^Z\equiv \left\langle Z^{\otimes n}\right\rangle.
\label{eq:gn_Z}
\end{equation}

Using Eqs.~\eqref{eq:xz-observable-general_app},
\eqref{eq:gn_X}, and \eqref{eq:gn_Z}, the correlators entering
Eq.~\eqref{eq:parity-chsh-general} take the form
\begin{equation}
\langle A_0B_y\rangle=u_{A_0}u_{B_y}G_2,
\label{eq:two-body-decomposition}
\end{equation}
and
\begin{align}
\left\langle
A_1B_y\prod_{j=1}^{n-2}C^{(j)}_1
\right\rangle
={}&
G_n^Z
u_{A_1}u_{B_y}
\prod_{j=1}^{n-2}u_{C^{(j)}_1}
\nonumber\\
&+G_n^X v_{A_1}v_{B_y}
\prod_{j=1}^{n-2}v_{C^{(j)}_1}.
\label{eq:n-body-decomposition}
\end{align}

\subsection{Exact optimization over the settings of party A and party B }
\label{app:measurement_optimization}
The observable $A_0$ appears only through $u_{A_0}$. Its magnitude
is therefore maximized by choosing $A_0=\pm Z$, i.e.,
$\lvert u_{A_0}\rvert=1$, with the sign chosen to make the
two-body contribution positive. Define
\begin{equation}
U\equiv |G_2|,
\quad
V\equiv G_n^Z u_{A_1}\prod_{j=1}^{n-2}u_{C^{(j)}_1},
\quad
W\equiv G_n^X v_{A_1}\prod_{j=1}^{n-2}v_{C^{(j)}_1}.
\label{eq:uvw-definitions}
\end{equation}
The Bell parameter then becomes
\begin{align}
S_n={}&(U+V)\,u_{B_0}+W\,v_{B_0}
\nonumber\\
&+(U-V)\,u_{B_1}-W\,v_{B_1}.
\label{eq:bell-before-bob-optimization}
\end{align}
The two vectors
\[(u_{B_y},v_{B_y}),\qquad y\in\{0,1\},\]
are independent unit vectors. Applying the Cauchy--Schwarz inequality
and maximizing the two scalar products independently gives
\begin{equation}
S_{\max}(U,V,W)=\sqrt{(U+V)^2+W^2}
+\sqrt{(U-V)^2+W^2}.
\label{eq:bell-max-reduced}
\end{equation}

For $0\leq U\leq1$ and $|V|\leq1$, squaring
Eq.~\eqref{eq:bell-max-reduced} gives
\begin{equation}
S_{\max}\leq2
\quad\Longleftrightarrow\quad
W^2\leq(1-U^2)(1-V^2).
\label{eq:no-violation-condition}
\end{equation}
This criterion allows us to establish the thresholds globally, without
assuming beforehand that the optimum lies in a particular measurement
sector.

For later use, define
\begin{equation}
C\equiv |u_{A_1}|\prod_{j=1}^{n-2}|u_{C^{(j)}_1}|,
\qquad
D\equiv |v_{A_1}|\prod_{j=1}^{n-2}|v_{C^{(j)}_1}|.
\label{eq:CD-definition}
\end{equation}
Since \(u_M^2+v_M^2=1\) for every observable,
\begin{equation}
C^2+D^2\leq1.
\label{eq:CD-bound}
\end{equation}

\subsection{Vacuum–$n$-photon state}

Applying the vacuum–$n$-photon entangled state $\ket{\Phi_n(q)}$ to Eqs.~\eqref{eq:gn_X} and~\eqref{eq:gn_Z} , the relevant
transfer functions are
\begin{align}
G_2^\phi&=
1-4\eta(1-\eta)q,
\label{eq:G2-phi}
\\
G_n^{Z,\phi}&=1-q\left[1-(1-2\eta)^n\right],
\label{eq:GnZ-phi}
\\
G_n^X&=2\eta^{n/2}\sqrt{q(1-q)}.
\label{eq:GnX-phi}
\end{align}
Introduce $a\equiv 4\eta(1-\eta)$, so that $U=1-aq$. Moreover,
\begin{equation}
|V|=|G_n^{Z,\phi}|C,
\qquad
|W|=2\eta^{n/2}\sqrt{q(1-q)}\,D.
\label{eq:VW-phi}
\end{equation}

We first prove that no violation is possible when
\begin{equation}
\eta^{n-1}\leq2(1-\eta).
\label{eq:phi-below-threshold}
\end{equation}
Equation~\eqref{eq:phi-below-threshold} implies
\begin{equation}
4\eta^n\leq8\eta(1-\eta)=2a.
\label{eq:phi-coherence-bound}
\end{equation}
Consequently,
\begin{equation}
W^2=4\eta^n q(1-q)D^2
\leq2a q(1-q)D^2.
\label{eq:phi-W-bound}
\end{equation}
On the other hand,
\begin{align}
(1-U^2)(1-V^2)&=aq(2-aq)
\left[1-(G_n^{Z,\phi})^2C^2\right]
\nonumber\\
&\geq aq(2-aq)(1-C^2)
\nonumber\\
&\geq2a q(1-q)D^2.
\label{eq:phi-UV-bound}
\end{align}
Here we used $|G_n^{Z,\phi}|\leq1$,
Eq.~\eqref{eq:CD-bound}, and
\begin{equation}
2-aq\geq2(1-q),
\end{equation}
which follows from $a=4\eta(1-\eta)\leq1$.
Combining Eqs.~\eqref{eq:phi-W-bound} and
\eqref{eq:phi-UV-bound} gives
\begin{equation}
W^2\leq(1-U^2)(1-V^2).
\end{equation}
Equation~\eqref{eq:no-violation-condition} therefore proves that
$S_n\leq2$ for every $q$ and every measurement setting whenever
Eq.~\eqref{eq:phi-below-threshold} holds.

It remains to show that a violation exists above this boundary. To prove achievability above this boundary, it is sufficient to exhibit one violating measurement choice. We choose
\begin{equation}
A_1=C^{(1)}_1=\cdots=C^{(n-2)}_1=X,
\label{eq:phi-threshold-settings}
\end{equation}
which eliminates the full $Z$-sector contribution and maximizes the coherence contribution. Equation \eqref{eq:bell-max-reduced} then reduces to
\begin{equation}
S_\phi(\eta,q)=
2\sqrt{\left[1-4\eta(1-\eta)q\right]^2
+4\eta^nq(1-q)}.
\label{eq:phi-S}
\end{equation}
Equivalently,
\begin{align}
\frac{S_\phi^2}{4}
={}&1+\left[4\eta^n-8\eta(1-\eta)\right]q
\nonumber\\
&+\left[16\eta^2(1-\eta)^2-4\eta^n\right]q^2.
\label{eq:phi-quadratic}
\end{align}

At $q=0$, the state is the product vacuum and $S_\phi=2$.
Moving away from $q=0$ produces a violation whenever the initial
slope of Eq.~\eqref{eq:phi-quadratic} is positive:
\begin{equation}
4\eta^n-8\eta(1-\eta)>0.
\end{equation}
For \(\eta>0\), this is equivalent to
\begin{equation} \label{eqn:cond:violation:phi}
\eta^{n-1}>2(1-\eta).
\end{equation}
Together with the no-violation result above, this proves that the exact
critical efficiency is the unique solution of
\begin{equation}
\boxed{
(\eta_{\text{th}}^\phi)^{n-1}=2(1-\eta_{\text{th}}^\phi).
}
\label{eq:phi-threshold}
\end{equation}

For $\eta>\eta_{\text{th}}^\phi$, the state weight that maximizes
Eq.~\eqref{eq:phi-S} is
\begin{equation}
q_{\text{opt}}^\phi(\eta)=\frac{
4\eta^n-8\eta(1-\eta)}{2\left[4\eta^n-16\eta^2(1-\eta)^2\right]},
\label{eq:phi-optimal-q}
\end{equation}
whenever the maximum lies in the interior of $q\in[0,1]$. In
particular,
\begin{equation}
q_{\text{opt}}^\phi(\eta)\longrightarrow0
\qquad
\text{as}
\qquad
\eta\longrightarrow(\eta_{\text{th}}^\phi)^+.
\end{equation}
The threshold is therefore approached by an increasingly weak entangled admixture on top of the loss-immune vacuum branch. At $\eta=\eta_{\text{th}}^\phi$, the limiting state has $q=0$ and only saturates the classical bound; a strict violation requires $\eta>\eta_{\text{th}}^\phi$.

\subsection{Fixed-photon-number state: complementary-occupation assignment}
\label{app:complementary-occupation-assignment}
For the fixed-photon-number state $\ket{\Psi_n(q)}$, with complementary-occupation assignment, the transfer functions are

\begin{align}
G_{2,\mathrm{comp}}^\psi
&=1-2\eta,
\label{eq:G2-psi}
\\
G_n^{Z,\psi}&=
(1-2\eta)^{n/2},
\label{eq:GnZ-psi}
\\
G_n^X&=
2\eta^{n/2}\sqrt{q(1-q)}.
\label{eq:GnX-psi}
\end{align}
Hence
\begin{equation}
U=|1-2\eta|,
\quad|V|=U^{n/2}C,
\quad|W|=2\eta^{n/2}\sqrt{q(1-q)}\,D.
\label{eq:UVW-psi}
\end{equation}

Because $4q(1-q)\leq1$,
\begin{equation}
W^2\leq\eta^nD^2.
\label{eq:psi-W-bound}
\end{equation}
Suppose that
\begin{equation}
\eta^{n-1}\leq4(1-\eta).
\label{eq:psi-below-threshold}
\end{equation}
Multiplying by \(\eta\) gives
\begin{equation}
\eta^n\leq4\eta(1-\eta)
=1-(1-2\eta)^2=1-U^2.
\label{eq:psi-coherence-bound}
\end{equation}
Moreover,
\begin{equation}
1-V^2=1-U^nC^2 \geq 1-C^2 \geq D^2.
\label{eq:psi-V-bound}
\end{equation}
It follows that
\begin{equation}
W^2\leq \eta^nD^2 \leq (1-U^2)(1-V^2).
\end{equation}
Therefore $S_n\leq2$ for all $q$ and all measurement settings whenever Eq.~\eqref{eq:psi-below-threshold} holds.

Above this boundary, choose
\begin{equation}
q=\frac12,
\qquad
A_1=C^{(1)}_1=\cdots=C^{(n-2)}_1=X.
\label{eq:psi-threshold-settings}
\end{equation}
Then \(V=0\) and \(W=\eta^{n/2}\), giving
\begin{equation}
S_\psi(\eta)=2\sqrt{(1-2\eta)^2+\eta^n}.
\label{eq:psi-S}
\end{equation}
A violation occurs when
\begin{equation} \label{eqn:cond:violation:psi:comp}
(1-2\eta)^2+\eta^n>1
\quad\Longleftrightarrow\quad
\eta^{n-1}>4(1-\eta).
\end{equation}
Together with the no-violation result above, this proves that the exact
critical efficiency is the unique solution of

\begin{equation}
\boxed{(\eta_{\text{th,comp}}^\psi)^{n-1}
=4(1-\eta_{\text{th,comp}}^\psi).}
\label{eq:psi-threshold}
\end{equation}

Unlike the vacuum-$n$-photon state, for the complementary-occupation assignment of $\ket{\Psi_n(q)}$, the optimal state remains maximally entangled,
\begin{equation}
q_{\text{opt,comp}}^\psi=\frac12,
\end{equation}
because its diagonal correlators are independent of $q$, while its coherence is maximized at $q=1/2$.

%==============================
\subsection{Fixed-photon-number state: equal-occupation assignment}
\label{app:equal-occupation-assignment}

We now assign the distinguished roles $A$ and $B$ to parties with
equal photon occupations in the two components of $\ket{\Psi_n(q)}$.
Such an assignment is available for every even $n\geq4$. Without loss
of generality, we choose the photon-number class for which the
occupations of $A$ and $B$ are $00$ in the component with weight
$1-q$ and $11$ in the component with weight $q$. Choosing the other
class is equivalent to the transformation $q\leftrightarrow1-q$.

Using Eq.~\eqref{eq:heisenberg-loss}, the two-body diagonal
correlator after loss is
\begin{equation}
G_{2,\mathrm{eq}}^\psi
=1-4\eta(1-\eta)q.
\label{eq:G2-psi-equal}
\end{equation}
This is identical to the two-body correlator $G_2^\phi$ in
Eq.~\eqref{eq:G2-phi}. By contrast, the complementary-occupation
assignment gives
$G_{2,\mathrm{comp}}^\psi=1-2\eta$.

The full diagonal and coherence correlators are unchanged relative
to the complementary-occupation assignment:
\begin{align}
G_n^{Z,\psi}
&=(1-2\eta)^{n/2},
\label{eq:GnZ-psi-equal}
\\
G_n^X
&=2\eta^{n/2}\sqrt{q(1-q)}.
\label{eq:GnX-psi-equal}
\end{align}

The no-violation proof for $\ket{\Phi_n(q)}$ applies without modification because it uses the same two-body correlator and coherence, together with the general bound
$|G_n^{Z,\psi}|\leq1$. For achievability, choosing
\begin{equation}
A_1=C^{(1)}_1=\cdots=C^{(n-2)}_1=X
\end{equation}
eliminates the full-$Z$ contribution. After optimizing party B's observables, the Bell parameter becomes
\begin{equation}
S_{\psi,\mathrm{eq}}(\eta,q)
=2\sqrt{\left[1-4\eta(1-\eta)q\right]^2+4\eta^nq(1-q)}.
\label{eq:psi-equal-S}
\end{equation}
This is identical to Eq.~\eqref{eq:phi-S}. The exact
equal-occupation threshold therefore coincides with that of the
vacuum--$n$-photon state:
\begin{equation}
\boxed{
\eta_{\mathrm{th,eq}}^\psi=\eta_{\mathrm{th}}^\phi,
\qquad
\left(\eta_{\mathrm{th,eq}}^\psi\right)^{n-1}=
2\left(1-\eta_{\mathrm{th,eq}}^\psi\right).
}
\label{eq:psi-equal-threshold}
\end{equation}

%==============================

\subsection{Full loss-density-matrix comparison}
\label{app:full-loss-density-matrix-comparison}
Here we derive the full density matrices obtained when the photon-loss channel acts on the two classes of states considered in the main text. The calculation includes all photon-loss contributions,
including those outside the original logical subspace.

For a single optical mode restricted to the $\{|0\rangle,|1\rangle\}$ subspace, photon loss is described by the amplitude-damping channel with Kraus operators given in Eq.~(\ref{eq:loss-kraus-threshold}), which satisfy $K_0^\dagger K_0+K_1^\dagger K_1=I$. For $n$ modes, the loss channel is
\begin{equation} \label{eqn:loss:ch}
	\mathcal{A}_{\eta}(\rho)
	= \sum_{\boldsymbol{\mu}\in\{0,1\}^n}
	K_{\boldsymbol{\mu}}
	\rho K_{\boldsymbol{\mu}}^\dagger,
\end{equation}
where
\begin{equation}
	K_{\boldsymbol{\mu}} =
	K_{\mu_1}\otimes\cdots\otimes K_{\mu_n}.
\end{equation}
The action on a computational basis is particularly simple. If a basis state contains $m$ photons and a particular set of $k$ photons survives, the corresponding probability is $\eta^k(1-\eta)^{m-k}$.

It is useful to introduce the notation $|1_S0_{\bar S}\rangle$ for the $n$-mode state in which the modes belonging to the subset $S\subseteq\{1,\ldots,n\}$ contain one photon and all other modes are in the vacuum. Thus, $|S|$ is the number of photons that survive.

Consider the vacuum-$n$-photon state, Eq.~(\ref{eq:phi_n}), with its corresponding density matrix $\rho_\phi \equiv  |\Phi_n\rangle\langle\Phi_n|$, and define effective logical states
\begin{equation}
	|0_L\rangle_{\phi}=|0\rangle^{\otimes n}, \qquad |1_L\rangle_{\phi}=|1\rangle^{\otimes n}.
\end{equation}
It is straightforward to show that by applying the loss channel to the logical-basis representation of the density matrix one obtains
\begin{equation}
	\rho_\phi(\eta) = \rho_{\phi,L} (\eta) + \tilde\rho_\phi (\eta),
\end{equation}
where 
\begin{equation} 
	\rho_{\phi,L}(\eta) = 
\begin{pmatrix}
	1-q+q(1-\eta)^n & \sqrt{q(1-q)}\,\eta^{n/2} \\
	\sqrt{q(1-q)}\,\eta^{n/2} & q\eta^n
\end{pmatrix},
\end{equation}
is the part of this state that remains within the original logical
subspace $	\mathcal{H}_{\phi} =\operatorname{span}\left\{|0_L\rangle_{\phi},|1_L\rangle_{\phi}\right\}$, and
\begin{equation}
	\tilde\rho_\phi (\eta) \equiv  q \sum_{k=1}^{n-1}
	\eta^k(1-\eta)^{n-k} \sum_{\substack{S\subseteq[n]\\ |S|=k}} |1_S0_{\bar S}\rangle \langle1_S0_{\bar S}|,
\end{equation}
denotes terms that do not belong to this subspace. Taking the trace of $\rho_{\phi,L}(\eta)$ gives the probability of remaining within the original
logical subspace,
\begin{equation} 	\label{eq:pL_phi_appendix}
	p_{\text L}^{(\phi)} = \operatorname{Tr}[\rho_{\phi,L}(\eta)] = 1-q+q[(1-\eta)^n+\eta^n].
\end{equation}
The two contributions proportional to $(1-\eta)^n$ and $\eta^n$ correspond, respectively, to complete loss and complete survival of the $n$-photon component. In particular, the complete-loss event maps the $n$-photon part onto the logical vacuum, but it does not preserve the coherence between the two logical states.

We now consider the fixed-photon state, where $\rho_\psi \equiv  |\Psi_n(q)\rangle \langle\Psi_n(q)|$, and define the corresponding logical states
\begin{equation}
	|0_L\rangle_{\psi} = |01\rangle^{\otimes n/2}, \qquad |1_L\rangle_{\psi}=|10\rangle^{\otimes n/2}.
\end{equation}
Further, let
\begin{equation}
	A=\{2,4,\ldots,n\}, \qquad B=\{1,3,\ldots,n-1\}
\end{equation}
denote the occupied modes in the two branches, respectively. Applying the loss channel (Eq.~(\ref{eqn:loss:ch})) to $\{|j_L\rangle_{\psi}\langle k_L|\}$ for $j,k\in\{0,1\}$ gives the full output density matrix, which can be decomposed into a contribution within the logical subspace and a contribution outside it:
\begin{equation}
	\rho_\psi(\eta) = \rho_{\psi,L} (\eta) + \tilde\rho_\psi (\eta),
\end{equation}
where
\begin{equation} 
	\rho_{\psi,L}(\eta) = \eta^{n/2}\, |\Psi_n(q)\rangle \langle\Psi_n(q)|,
\end{equation}
represents the part that remains in the original logical subspace $	\mathcal{H}_{\psi} =\operatorname{span}\left\{|0_L\rangle_{\psi},|1_L\rangle_{\psi}\right\}$, and
\begin{align}
	\tilde\rho_\psi (\eta) =& (1-q) \sum_{\substack{S\subseteq A \\ |S|\neq n/2}} \eta^{|S|} (1-\eta)^{n/2-|S|}
	|1_S\rangle\langle1_S| + \nonumber \\
    &\quad q \sum_{\substack{T\subseteq B \\ |T|\neq n/2}} \eta^{|T|} (1-\eta)^{n/2-|T|}
	|1_T\rangle\langle1_T|,
\end{align}
is part of the state outside of the logical subspace. Here $|1_{S/T}\rangle$ means the $n$-mode state with photons exactly in the mode $S/T$. Note that both logical states remain in this subspace only when all $n/2$ photons survive. Thus, $p_{\text L}^{(\psi)} = \operatorname{Tr}[\rho_{\psi,L}(\eta)] = \eta^{n/2}$.

The density-matrix decomposition is independent of which photon-number class is assigned to the distinguished parties $A$ and $B$. The equal- and complementary-occupation assignments introduced in the main text therefore do not change the total logical-subspace survival probability. Their difference instead appears in the particular two-body correlator $G_2$ entering the parity--CHSH functional.

\subsection{Physical origin of the different loss tolerances}

The difference between the detection-efficiency thresholds under the complementary-occupation assignment follows from the way photon loss modifies the correlators entering the parity--CHSH functional. In both state families, the coherence correlation has the same loss dependence, Eqs.~(\ref{eq:GnX-psi}), (\ref{eq:GnX-psi-equal}), and (\ref{eq:GnX-phi}),
so the lower threshold of the vacuum--$n$-photon state is not due to more favourable preservation of its off-diagonal coherence. Instead, the distinction arises from the two-body diagonal correlator. For the vacuum--$n$-photon state, $G_2^\phi$
 can be made arbitrarily close to its loss-free value by taking $q\rightarrow0$, while retaining nonzero coherence for any $q>0$. In contrast, under the complementary-occupation assignment for $\ket{\Psi_n(q)}$, we obtain Eq.~(\ref{eq:G2-psi})
which is independent of $q$. Consequently, state optimization can compensate for the loss-induced reduction of the diagonal correlator in the vacuum--$n$-photon encoding but not in the complementary fixed-photon-number encoding. This difference, together with the nonlinear form of the optimized parity--CHSH functional, leads to the respective conditions in Eqs.~(\ref{eqn:cond:violation:phi}) and (\ref{eqn:cond:violation:psi:comp}).
For the equal-occupation assignment, however, the fixed-photon-number state has
$G_{2,\mathrm{eq}}^\psi$ as given in Eq.~(\ref{eq:G2-psi-equal}), and therefore recovers the same threshold as the vacuum--$n$-photon state, as shown above.

A complementary physical picture follows from the full loss-channel density matrices derived above. The vacuum-$n$-photon encoding contains a logical vacuum component, $|0_L\rangle_{\phi}=|0\rangle^{\otimes n}$,
which is invariant under photon loss, whereas both logical states $|0_L\rangle_{\psi}=|01\rangle^{\otimes n/2}$ and $|1_L\rangle_{\psi}=|10\rangle^{\otimes n/2}$
of the fixed-photon-number encoding contain $n/2$ photons and are therefore subject to loss. The corresponding weights remaining in the original logical subspaces are shown to be
\begin{equation}
	p_{\text{L}_q}^{(\phi)} = 1-q+q[(1-\eta)^n+\eta^n],
	\qquad
	p_{\text{L}_q}^{(\psi)} = \eta^{n/2}.
\end{equation}
These probabilities provide useful intuition for the different response of the two encodings to loss, but they should not be
interpreted as measures of coherence preservation. In the vacuum-$n$-photon encoding, the vacuum component is loss invariant,
while the all-photons-lost event from the $n$-photon component also returns the state to the logical vacuum. The latter event, however, does not preserve the coherence between the two logical states. In the fixed-photon-number encoding, no loss event other than the no-loss event remains within the original logical subspace: every logical branch contains $n/2$ photons, and all $n/2$ photons must survive. It is instructive to compare the logical-subspace probabilities in a few limiting cases. In the limit $q\rightarrow0$, one obtains
\begin{equation}
	p_{\text L_0}^{(\phi)}\longrightarrow 1,
	\qquad
	p_{\text L_0}^{(\psi)}=\eta^{n/2}.
\end{equation}
Thus, when the vacuum state dominates, the vacuum-$n$-photon encoding remains entirely within its logical subspace in this limit, whereas the fixed-photon-number encoding retains only the probability $\eta^{n/2}$ of remaining in its original logical subspace.

For a balanced superposition, $q=1/2$, the corresponding probabilities are
\begin{equation}
		p_{\text L_{1/2}}^{(\phi)}=\frac{1}{2}[1+\eta^n+(1-\eta)^n], \qquad p_{\text L_{1/2}}^{(\psi)}=\eta^{n/2}.
\end{equation}
Since $0\leq\eta\leq1$ and $n>0$, these satisfy
\begin{equation}
    p_{\text L_{1/2}}^{(\phi)} - p_{\text L_{1/2}}^{(\psi)} = \frac{1}{2}\left[(1-\eta)^n + (1-\eta^{n/2})^2\right] \ge 0,
\end{equation}
 with equality only for $\eta=1$ (or trivially at $n=0$). Hence, at balanced weighting, the vacuum-$n$-photon encoding has a greater probability of remaining in its original logical subspace.
These limiting cases provide useful information about the robustness of the two encodings under loss. However, the logical-subspace probabilities alone do not determine the Bell-violation threshold. In particular, remaining in the logical subspace does not necessarily mean that the original logical coherence has survived. The Bell threshold is instead governed by the full loss-affected correlations entering the parity--CHSH expression. Thus, the decisive difference for parity--CHSH violation remains the behavior of the diagonal correlator $G_2$, together with the particular way in which $G_2$,
$G_n^Z$, and $G_n^X$ enter the Bell functional.

%====================
\section{Ideal Heralding Scheme}
\subsection{Ideal Initial State Expansion}
\label{app:expansion}

Expanding Eq.~(\ref{eq:global_input}), we obtain 16 terms. We write the state in the form
\[
\ket{\Psi_{\text{in}}} =\sum_ic_i\ket{\text{retained}_i}_{a_1 b_1 a_2 b_2} \otimes \ket{\text{transmitted}_i}_{a_1' b_1' a_2' b_2'}.
\]

The complete expansion is:
\begin{align}
\ket{\Psi_{\text{in}}} &= 
\alpha_1\gamma_1\alpha_2\gamma_2 \ket{1010} \otimes \ket{0000} \nonumber\\
&+ \alpha_1\gamma_1\alpha_2\delta_2 \ket{1011} \otimes \ket{0001} \nonumber\\
&+ \alpha_1\gamma_1\beta_2\gamma_2 \ket{1000} \otimes \ket{0010} \nonumber\\
&+ \alpha_1\gamma_1\beta_2\delta_2 \ket{1001} \otimes \ket{0011} \nonumber\\
&+ \alpha_1\delta_1\alpha_2\gamma_2 \ket{1110} \otimes \ket{0100} \nonumber\\
&+ \alpha_1\delta_1\alpha_2\delta_2 \ket{1111} \otimes \ket{0101} \nonumber\\
&+ \alpha_1\delta_1\beta_2\gamma_2 \ket{1100} \otimes \ket{0110} \nonumber\\
&+ \alpha_1\delta_1\beta_2\delta_2 \ket{1101} \otimes \ket{0111} \nonumber\\
&+ \beta_1\gamma_1\alpha_2\gamma_2 \ket{0010} \otimes \ket{1000} \nonumber\\
&+ \beta_1\gamma_1\alpha_2\delta_2 \ket{0011} \otimes \ket{1001} \nonumber\\
&+ \beta_1\gamma_1\beta_2\gamma_2 \ket{0000} \otimes \ket{1010} \nonumber\\
&+ \beta_1\gamma_1\beta_2\delta_2 \ket{0001} \otimes \ket{1011} \nonumber\\
&+ \beta_1\delta_1\alpha_2\gamma_2 \ket{0110} \otimes \ket{1100} \nonumber\\
&+ \beta_1\delta_1\alpha_2\delta_2 \ket{0111} \otimes \ket{1101} \nonumber\\
&+ \beta_1\delta_1\beta_2\gamma_2 \ket{0100} \otimes \ket{1110} \nonumber\\
&+ \beta_1\delta_1\beta_2\delta_2 \ket{0101} \otimes \ket{1111},
\label{eq:full_expansion}
\end{align}
where mode subscripts have been suppressed for clarity.

\Cref{tab:state_expansion} organizes these terms by the total photon number $n$ in the transmitted modes.

\begin{table}[h]
\centering
\caption{Complete expansion of the input state $\ket{\Psi_{\text{in}}}$ organized by total photon number $n$ in the transmitted modes. The mode ordering is $(a_1, b_1, a_2, b_2)$ for retained modes and $(a_1', b_1', a_2', b_2')$ for transmitted modes. Only the $n=2$ terms contribute to successful heralding events with PNR detectors requiring exactly two coincident clicks.}
\label{tab:state_expansion}
\begin{tabular}{cccc}
\toprule
$n$ & Coefficient & Retained & Transmitted \\
\midrule
0 & $\alpha_1\gamma_1\alpha_2\gamma_2$ & $\ket{1010}$ & $\ket{0000}$ \\
\midrule
\multirow{4}{*}{1} & $\alpha_1\gamma_1\alpha_2\delta_2$ & $\ket{1011}$ & $\ket{0001}$ \\
& $\alpha_1\gamma_1\beta_2\gamma_2$ & $\ket{1000}$ & $\ket{0010}$ \\
& $\alpha_1\delta_1\alpha_2\gamma_2$ & $\ket{1110}$ & $\ket{0100}$ \\
& $\beta_1\gamma_1\alpha_2\gamma_2$ & $\ket{0010}$ & $\ket{1000}$ \\
\midrule
\multirow{6}{*}{2} & $\alpha_1\gamma_1\beta_2\delta_2$ & $\ket{1001}$ & $\ket{0011}$ \\
& $\alpha_1\delta_1\alpha_2\delta_2$ & $\ket{1111}$ & $\ket{0101}$ \\
& $\alpha_1\delta_1\beta_2\gamma_2$ & $\ket{1100}$ & $\ket{0110}$ \\
& $\beta_1\gamma_1\alpha_2\delta_2$ & $\ket{0011}$ & $\ket{1001}$ \\
& $\beta_1\gamma_1\beta_2\gamma_2$ & $\ket{0000}$ & $\ket{1010}$ \\
& $\beta_1\delta_1\alpha_2\gamma_2$ & $\ket{0110}$ & $\ket{1100}$ \\
\midrule
\multirow{4}{*}{3} & $\alpha_1\delta_1\beta_2\delta_2$ & $\ket{1101}$ & $\ket{0111}$ \\
& $\beta_1\gamma_1\beta_2\delta_2$ & $\ket{0001}$ & $\ket{1011}$ \\
& $\beta_1\delta_1\alpha_2\delta_2$ & $\ket{0111}$ & $\ket{1101}$ \\
& $\beta_1\delta_1\beta_2\gamma_2$ & $\ket{0100}$ & $\ket{1110}$ \\
\midrule
4 & $\beta_1\delta_1\beta_2\delta_2$ & $\ket{0101}$ & $\ket{1111}$ \\
\bottomrule
\end{tabular}
\end{table}

%====================================
\subsection{Central-Node Interferometer}
\label{app:interferometer}

The transmitted modes undergo interference at a central station via a linear-optical network. We employ a two-layer beam-splitter configuration (\Cref{fig:Heralding_scheme}), following the standard four-mode convention used in the literature.

\textit{First layer}: 50:50 beam splitters couple modes separated by two positions:
\begin{align}
(a_1', a_2') &\xrightarrow{\text{BS}_1} (c_1, c_3), \\
(b_1', b_2') &\xrightarrow{\text{BS}_2} (c_2, c_4).
\end{align}

\textit{Second layer}: 50:50 beam splitters couple the intermediate modes within each pair:
\begin{align}
(c_1, c_2) &\xrightarrow{\text{BS}_3} (d_1, d_2), \\
(c_3, c_4) &\xrightarrow{\text{BS}_4} (d_3, d_4).
\end{align}

The output modes $d_1, d_2, d_3, d_4$ are monitored by single-photon detectors $D_1, D_2, D_3, D_4$.

For a 50:50 beam splitter with the convention
\begin{equation}
U_{\text{BS}} = \frac{1}{\sqrt{2}}
\begin{pmatrix}
1 & 1 \\
1 & -1
\end{pmatrix},
\end{equation}
the complete transformation from input to detector modes is computed by composing the two layers.

We derive the transformation between input-mode and detector-mode creation operators by applying the beam-splitter transformations sequentially.

\textit{First layer}:
\begin{align}
c_1^\dagger &= \tfrac{1}{\sqrt{2}}(a_1'^\dagger + a_2'^\dagger), &
c_3^\dagger &= \tfrac{1}{\sqrt{2}}(a_1'^\dagger - a_2'^\dagger), \\
c_2^\dagger &= \tfrac{1}{\sqrt{2}}(b_1'^\dagger + b_2'^\dagger), &
c_4^\dagger &= \tfrac{1}{\sqrt{2}}(b_1'^\dagger - b_2'^\dagger).
\end{align}

\textit{Second layer}:
\begin{align}
d_1^\dagger &= \tfrac{1}{\sqrt{2}}(c_1^\dagger + c_2^\dagger), &
d_2^\dagger &= \tfrac{1}{\sqrt{2}}(c_1^\dagger - c_2^\dagger), \\
d_3^\dagger &= \tfrac{1}{\sqrt{2}}(c_3^\dagger + c_4^\dagger), &
d_4^\dagger &= \tfrac{1}{\sqrt{2}}(c_3^\dagger - c_4^\dagger).
\end{align}

Composing these transformations, the detector-mode operators in terms of input-mode operators are:
\begin{align}
d_1^\dagger &= \tfrac{1}{2}(a_1'^\dagger + b_1'^\dagger + a_2'^\dagger + b_2'^\dagger), \\
d_2^\dagger &= \tfrac{1}{2}(a_1'^\dagger - b_1'^\dagger + a_2'^\dagger - b_2'^\dagger), \\
d_3^\dagger &= \tfrac{1}{2}(a_1'^\dagger + b_1'^\dagger - a_2'^\dagger - b_2'^\dagger), \\
d_4^\dagger &= \tfrac{1}{2}(a_1'^\dagger - b_1'^\dagger - a_2'^\dagger + b_2'^\dagger).
\end{align}

The corresponding unitary matrix, with input ordering $(a_1', b_1', a_2', b_2')$ and output ordering $(d_1,d_2,d_3,d_4)$, is
\begin{equation}
U = \frac{1}{2}
\begin{pmatrix}
1 & 1 & 1 & 1 \\
1 & -1 & 1 & -1 \\
1 & 1 & -1 & -1 \\
1 & -1 & -1 & 1
\end{pmatrix}.
\label{eq:U_matrix_appendix}
\end{equation}
This matrix is orthogonal and symmetric, satisfying $U^2=I$. Consequently, $U^{-1}=U$, and the input-mode operators expressed in terms of detector-mode operators are:
\begin{align}
a_1'^\dagger &= \tfrac{1}{2}(d_1^\dagger + d_2^\dagger + d_3^\dagger + d_4^\dagger), \label{eq:a1_transform} \\
b_1'^\dagger &= \tfrac{1}{2}(d_1^\dagger - d_2^\dagger + d_3^\dagger - d_4^\dagger), \label{eq:b1_transform} \\
a_2'^\dagger &= \tfrac{1}{2}(d_1^\dagger + d_2^\dagger - d_3^\dagger - d_4^\dagger), \label{eq:a2_transform} \\
b_2'^\dagger &= \tfrac{1}{2}(d_1^\dagger - d_2^\dagger - d_3^\dagger + d_4^\dagger). \label{eq:b2_transform}
\end{align}
%====================================
\subsection{Projection onto the Two-Click Subspace}
\label{app:projectors}

For a given two-click detection pattern $(D_n,D_m)$, the action of the interferometer and detectors is equivalent to a rank-one measurement on the transmitted modes. Within the bunched-free two-photon subspace populated by the input state, the corresponding measurement operator is
\begin{equation}
\Pi_{nm}=
\ket{\chi_{nm}}\!\bra{\chi_{nm}},
\qquad
1\leq n<m\leq4,
\label{eq:projector_definition}
\end{equation}
where $\ket{\chi_{nm}}$ is the effective state associated with the detection pattern $(D_n,D_m)$ upon which the incoming state is projected.

Let $\hat{\mathcal U}$ denote the Fock-space unitary associated with the mode-transformation matrix $U$. The measurement ket $\ket{\chi_{nm}}$ is obtained by back-propagating the corresponding output state through the interferometer and retaining only the components containing at most one photon in each input mode. This restriction is sufficient because the transmitted component $\ket{\Psi_{2\text{ph}}}$ contains no terms with two photons occupying the same mode.

Since the interferometer matrix is both unitary and Hermitian, $U^{-1}=U$, and the effective measurement kets associated with the six two-mode coincidence outcomes are
\begin{subequations}
\label{eq:all_projectors}
\begin{align}
\ket{1100}_{d}
&\xrightarrow{\hat{\mathcal U}^{\dagger}}
\ket{\chi_{12}}
=\frac{1}{2}\left(\ket{1010}-\ket{0101}\right),
\label{eq:proj12}
\\
\ket{1010}_{d}
&\xrightarrow{\hat{\mathcal U}^{\dagger}}
\ket{\chi_{13}}=
\frac{1}{2}\left(\ket{1100}-\ket{0011}\right),
\label{eq:proj13}
\\
\ket{1001}_{d}
&\xrightarrow{\hat{\mathcal U}^{\dagger}}
\ket{\chi_{14}}
=\frac{1}{2}\left(\ket{1001}-\ket{0110}\right),
\label{eq:proj14}
\\
\ket{0110}_{d}
&\xrightarrow{\hat{\mathcal U}^{\dagger}}
\ket{\chi_{23}}
=\frac{1}{2}\left(\ket{0110}-\ket{1001}\right),
\label{eq:proj23}
\\
\ket{0101}_{d}
&\xrightarrow{\hat{\mathcal U}^{\dagger}}
\ket{\chi_{24}}
=
\frac{1}{2}\left(\ket{0011}-\ket{1100}\right),
\label{eq:proj24}
\\
\ket{0011}_{d}
&\xrightarrow{\hat{\mathcal U}^{\dagger}}
\ket{\chi_{34}}
=\frac{1}{2}
\left(\ket{0101}-\ket{1010}\right).
\label{eq:proj34}
\end{align}
\end{subequations}
Here, the states on the right-hand sides are expressed in the input-mode basis $(a_1',b_1',a_2',b_2')$. The effective measurement kets are unnormalized because components containing two photons in the same input mode have been omitted; such components are orthogonal to the input state and do not contribute to the heralding probabilities or conditional states.

The measurement kets occur in pairs that differ only by a global phase:
\begin{align}
\ket{\chi_{12}}&=-\ket{\chi_{34}},
\\
\ket{\chi_{13}}&=-\ket{\chi_{24}},
\\
\ket{\chi_{14}}&=-\ket{\chi_{23}}.
\end{align}
This symmetry reflects the structure of the interferometer and ensures that complementary detection patterns herald states differing only by a global phase, consistent with Table~\ref{tab:2click_heralded_states}.

%============================-
\subsection{Extension of the Heralded GHZ Distribution Protocol to More Than Four Parties}\label{app:extension}
In the main text, we focus on the case where $n = 4$. However, we can extend the heralded GHZ distribution protocol to more than four parties in a simple manner. In this appendix, we describe how to distribute the vacuum-$n$-photon GHZ state among six remote parties efficiently. Extension to more than six parties is straightforward.

\begin{figure}[!t]
 \centering
 \includegraphics[width=0.8\linewidth]{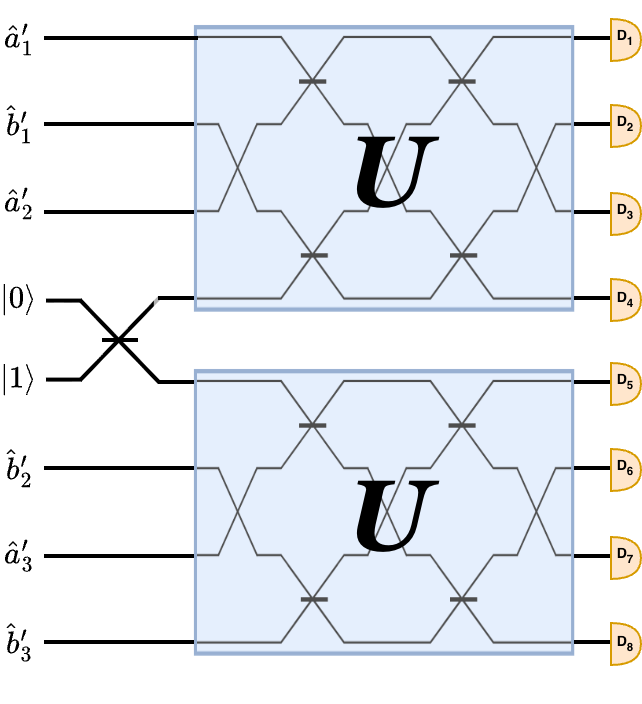}
 \caption{Schematic of the efficient heralded distribution of GHZ states for six parties.}
 \label{fig:6party}
\end{figure}

Figure~\ref{fig:6party} shows a schematic of the protocol for $n = 6$. Three of the six parties possess the Type-A entanglement sources, and the remaining three parties possess the Type-B sources. At the central station, we have two copies of the beam-splitter network $U$. We consider events where the detectors $D_1, D_2, D_5$, and $D_6$ each register one photon. Using the interferometer transformation of $U$ in Eq.~(\ref{eq:U_matrix_appendix}), the corresponding effective measurement kets on the upper and lower interferometer inputs are
\begin{align}
    \ket{\chi_{12}} &= \frac{1}{2}(\ket{1010}-\ket{0101})_{a_1'b_1'a_2' r_1},\\
    \ket{\chi_{56}} &= \frac{1}{2}(\ket{1010}-\ket{0101})_{r_2b_2'a_3'b_3'},
\end{align}
respectively, where $r_1$ and $r_2$ represent ancilla modes between the two interferometers. We prepare the ancilla modes $(r_1, r_2)$ in the following SPPE state $\ket{\psi_\text{SPPE}}$
\begin{equation}
    \ket{\psi_\text{SPPE}} = \frac{1}{\sqrt{2}} (\ket{01} - \ket{10}).
\end{equation}
Therefore, the resulting unnormalized effective measurement ket on the six transmitted modes is
\begin{equation}
    \begin{split}
        \ket{\chi_\text{eff}} &= \bra{\psi_\text{SPPE}} (\ket{\chi_{12}}\otimes\ket{\chi_{56}})\\
        &= \frac{1}{4\sqrt{2}} (\ket{101010} - \ket{010101}),
    \end{split}
\end{equation}
where the mode ordering is $(a_1', b_1', a_2', b_2', a_3', b_3')$. For the source states prepared by the six parties, this heralding event produces the following unnormalized state of the retained modes
\begin{equation}
\beta_1\gamma_1\beta_2\gamma_2\beta_3\gamma_3 \ket{000000} - \alpha_1\delta_1\alpha_2\delta_2\alpha_3\delta_3 \ket{111111}
\end{equation}
up to normalization. While we focus on the vacuum-$n$-photon GHZ state, we can distribute GHZ states of other forms by considering other detection patterns at the central station. We can straightforwardly extend this protocol to more than six parties. When the number of remote parties is $n$, we require $n/2-1$ copies of the beam-splitter network~\cite{shimizu2025simple}.

%==================================
\section{Heralding Performance under Photon Loss}
\label{app:psucc_and_fidelity_under_loss_app}

Here, we derive the heralding probabilities and conditional-state fidelities in the presence of photon loss. We treat separately the detection patterns associated with the target vacuum-$n$-photon GHZ state and those associated with fixed-photon-number GHZ-like states. We assume that the sources are identical within each type:
$\alpha_1=\alpha_2\equiv \alpha$,
$\beta_1=\beta_2\equiv \beta$,
$\gamma_1=\gamma_2\equiv \gamma$, and
$\delta_1=\delta_2\equiv \delta$.

Photon loss in each link is modelled by coupling every transmitted mode $x_i'$, where $x\in\{a,b\}$ and $i\in\{1,2\}$, to an initially unoccupied environment mode $E_{x_i}$ through the beam-splitter transformation
\begin{equation}
\ket{1}_{x_i'}\ket{0}_{E_{x_i}}
\longmapsto
\sqrt{\eta_{\text{c}}}\,
\ket{1}_{x_i'}\ket{0}_{E_{x_i}}
+
\sqrt{1-\eta_{\text{c}}}\,
\ket{0}_{x_i'}\ket{1}_{E_{x_i}},
\label{eq:loss_bs_app}
\end{equation}
followed by tracing over the environment modes. Here, $\eta_{\text c}$ denotes the effective transmissivity from each source to an ideal detector at the central station and may include the heralding-detector efficiency. We condition on events in which exactly two distinct detectors each register one photon, while the remaining two detectors register no photons.

\subsection{Target vacuum-$n$-photon GHZ State}
We first consider the detection pattern $(D_1,D_2)$. The analysis of $(D_3,D_4)$ is identical, and these are the two patterns associated with the target state, as listed in Table~\ref{tab:2click_heralded_states}. According to Eq.~\eqref{eq:d1d2_projector}, a  coincidence at $(D_1,D_2)$ can originate only from a photon pair in the transmitted modes $\{a_1',a_2'\}$ or $\{b_1',b_2'\}$. Each such pair produces the selected coincidence with probability $1/4$. The components of the input state Eq.~\eqref{eq:global_input} with a finite probability of producing this pattern therefore fall into two families: (i) both Type-A photons are transmitted and detected, while $m$ Type-B pairs, with $m\in\{0,1,2\}$, are emitted and their transmitted photons are lost; (ii) both Type-B pairs are emitted and their transmitted photons are detected, while $m$ Type-A photons are transmitted and lost. The probability amplitudes of these components are
\begin{equation}
 Y_m^{(\text{i})}=\beta^2\gamma^{2-m}\delta^m,
 \qquad
 Y_m^{(\text{ii})}=\delta^2\alpha^{2-m}\beta^m.
 \label{eq:Ym}
\end{equation}
with $\binom{2}{m}$ terms for each $m$ in each family, and the probability that the two designated photons are detected at $(D_1,D_2)$ while all $m$ additional transmitted photons are lost is
\begin{equation}
 p_m=\frac{\eta_{\text{c}}^2(1-\eta_{\text{c}})^m}{4}.
 \label{eq:pm}
\end{equation}
The probability of obtaining this detection pattern is therefore
\begin{widetext}
\begin{equation}
\begin{aligned}
P^{\Phi}_{\text{succ}}&=\sum_{m=0}^{2}\binom{2}{m}p_m
 \left(
   \left|Y_m^{(\text{i})}\right|^2
  +\left|Y_m^{(\text{ii})}\right|^2
 \right)\\
 &=\frac{\eta_{\text{c}}^2}{4}|\beta|^4
 \sum_{m=0}^{2}\binom{2}{m}
 (|\gamma|^2)^{2-m}
 \bigl(|\delta|^2(1-\eta_{\text{c}})\bigr)^m
 +\frac{\eta_{\text{c}}^2}{4}|\delta|^4
 \sum_{m=0}^{2}\binom{2}{m}
 (|\alpha|^2)^{2-m}
 \bigl(|\beta|^2(1-\eta_{\text{c}})\bigr)^m\\
 &=\frac{\eta_{\text{c}}^2}{4}
 \Bigl[
 |\beta|^4\bigl(|\gamma|^2+|\delta|^2(1-\eta_{\text{c}})\bigr)^2+|\delta|^4\bigl(|\alpha|^2+|\beta|^2(1-\eta_{\text{c}})\bigr)^2
 \Bigr].
 \label{eq:Psucc_pattern_app}
\end{aligned}
\end{equation}
\end{widetext}

Accepting both $(D_1,D_2)$ and $(D_3,D_4)$ gives the total raw probability associated with heralding the target state,
\begin{equation}
 P^{\Phi}_{\text{tot}}=2\times P^{\Phi}_{\text{succ}}.
 \label{eq:target_tot_app}
\end{equation}
The $m=0$ contribution represents genuine heralding events, whereas the $m=1$ and $m=2$ contributions represent loss-induced false heralds. In the lossless limit, only the genuine contribution remains.

For the symmetric family
$|\alpha|^2=|\gamma|^2\equiv  q$, the normalization conditions imply
$|\beta|^2=|\delta|^2=1-q$, and the total target-state heralding probability becomes
\begin{equation}
P^{\Phi}_{\text{tot}}
=
\eta_{\text{c}}^2(1-q)^2
\left[
q+(1-q)(1-\eta_{\text{c}})
\right]^2.
\label{eq:P_balanced_app}
\end{equation}

For the pattern $(D_1,D_2)$, collecting the contributions described above gives the normalized conditional state
\begin{widetext}
\begin{equation}
\begin{aligned}
 \rho_\Phi=\frac{1}{\mathcal{N}}
\Bigl[p_0\ket{\widetilde\Phi}\!\bra{\widetilde\Phi}+p_1\left|Y_1^{(\text{i})}\right|^2
\bigl(\ket{0100}\!\bra{0100}+\ket{0001}\!\bra{0001}\bigr) &+p_1\left|Y_1^{(\text{ii})}\right|^2
\bigl(\ket{0111}\!\bra{0111}+\ket{1101}\!\bra{1101}\bigr)
 \\
 &+p_2
 \left(
  \left|Y_2^{(\text{i})}\right|^2
 +\left|Y_2^{(\text{ii})}\right|^2
 \right)
 \ket{0101}\!\bra{0101}
 \,\Bigr],
\end{aligned}
\label{eq:rhoX}
\end{equation}
\end{widetext}
where $\ket{\widetilde{\Phi}}=\beta^2\gamma^2\ket{0000}
-\alpha^2\delta^2\ket{1111}$
is the desired unnormalized GHZ component and
$\mathcal{N}=P^{\Phi}_{\text{succ}}$ is the normalization factor.

The two histories where no photons were lost remain coherent. On the other hand, histories involving one or two lost photons leave distinguishable environment records and therefore contribute incoherently after the environment modes are traced out. These false-heralding components are orthogonal to the desired GHZ subspace.

After correcting the known relative phase, the fidelity with the target state $\ket{\text{GHZ}_4}=\ket{\Phi_4(0.5)}$ is
\begin{equation}
\mathcal F_{\Phi}
\equiv 
\bra{\text{GHZ}_4}
\rho_\Phi
\ket{\text{GHZ}_4}.
\label{eq:fidelity_definition_app}
\end{equation}
For arbitrary complex source amplitudes, this gives
\begin{equation}
\mathcal F_{\Phi}
=
\frac{
\left(
|\beta|^2|\gamma|^2
+
|\alpha|^2|\delta|^2
\right)^2
}{
2\mathcal D
},
\label{eq:fidelity_app}
\end{equation}
where
\begin{equation}
\mathcal D
=
|\beta|^4
\bigl(
|\gamma|^2+|\delta|^2(1-\eta_{\text{c}})
\bigr)^2
+
|\delta|^4
\bigl(
|\alpha|^2+|\beta|^2(1-\eta_{\text{c}})
\bigr)^2.
\label{eq:fidelity_denominator_app}
\end{equation}
Equivalently, since
$P^{\Phi}_{\text{succ}}=\eta_{\text c}^2\mathcal D/4$,
\begin{equation}
\mathcal F_{\Phi}
=
\frac{
\eta_{\text c}^2
\left(
|\beta|^2|\gamma|^2
+
|\alpha|^2|\delta|^2
\right)^2
}{
8P^{\Phi}_{\text{succ}}
}.
\label{eq:fidelity_psucc_app}
\end{equation}

For the symmetric family, the fidelity reduces to
\begin{equation}
\mathcal F_{\Phi}
=
\left[
\frac{q}
{q+(1-q)(1-\eta_{\text{c}})}
\right]^2.
\label{eq:F_balanced_app}
\end{equation}

The raw target-state heralding probability is maximized at
\begin{equation}
q_{\text{opt}}(\eta_{\text c})=
1-\frac{1}{2\eta_{\text c}},
\qquad
\eta_{\text c}>\frac{1}{2}.
\label{eq:popt_app}
\end{equation}
For $\eta_{\text c}\leq1/2$, it decreases monotonically with $q$ and is maximized at the boundary $q=0$. At this boundary, the genuine GHZ contribution vanishes and all accepted events are false heralds. In the lossless limit, the rate-optimal operating point approaches $q_{\text{opt}}=1/2$.

For the symmetric family, the projector-based GHZ witness certifies genuine multipartite entanglement whenever $\mathcal F_{\Phi}>\frac{1}{2}$,
or equivalently,
\begin{equation}
q>\frac{1-\eta_{\text c}}
{\sqrt{2}-\eta_{\text c}}.
\label{eq:q_entanglement_threshold_app}
\end{equation}
Moreover, $\lim_{\eta_{\text c}\to0}
\mathcal F_{\Phi}=q^2$.
Thus, $q>1/\sqrt{2}$ is sufficient to certify genuine multipartite entanglement at arbitrarily long distances within this model. For $\eta_{\text c}>1/2$, the fidelity at the raw-rate-optimal point is
\begin{equation}
\mathcal F_{\Phi}
\bigl(q_{\text{opt}}(\eta_{\text c})\bigr)
=\left(2-\frac{1}{\eta_{\text c}}
\right)^2.
\label{eq:F_at_qopt_app}
\end{equation}
This exceeds the biseparable bound when
\begin{equation}
\eta_{\text c}>\frac{1}{2-1/\sqrt{2}}
=\frac{4+\sqrt{2}}{7}\approx0.7735.
\label{eq:eta_entanglement_threshold_app}
\end{equation}
Below this transmissivity, certifying genuine multipartite entanglement requires choosing $q>q_{\text{opt}}$ and accepting a lower target-state heralding probability.

\subsection{Fixed-Photon-Number GHZ-Like States}
We next consider the four detection patterns $(D_1,D_3)$, $(D_2,D_4)$, $(D_1,D_4)$, and $(D_2,D_3)$. For example, a coincidence at $(D_1,D_3)$ can arise from a photon pair in either $\{a_1',b_1'\}$ or $\{a_2',b_2'\}$. The remaining Type-A photon is either retained locally, with amplitude $\alpha$, or transmitted and lost, with amplitude $\beta$. Similarly, the remaining Type-B source either emits no pair, with amplitude $\gamma$, or emits a pair whose transmitted photon is lost, with amplitude $\delta$.

Let $m_A,m_B\in\{0,1\}$ denote the numbers of additional lost photons originating from the Type-A and Type-B sources, respectively. The corresponding source amplitude and detection probability are
\begin{align}
Y_{m_A m_B}
&=\beta\delta\,
\alpha^{1-m_A}\beta^{m_A}
\gamma^{1-m_B}\delta^{m_B},
\\
p_{m_A m_B}&=\frac{\eta_{\text c}^2
(1-\eta_{\text c})^{m_A+m_B}}{4}.
\label{eq:Ym_mixed_app}
\end{align}
For each $(m_A,m_B)$, there is one contribution from each of the two input photon pairs that can produce the specified detection pattern. The probability of one specified pattern is therefore
\begin{align}
P^{\Psi}_{\text{succ}}
&=2\sum_{m_A,m_B}p_{m_A m_B}\left|Y_{m_A m_B}\right|^2
\nonumber\\
&=\frac{\eta_{\text c}^2}{2}|\beta|^2|\delta|^2
\bigl(|\alpha|^2+|\beta|^2(1-\eta_{\text c})
\bigr)
\bigl(|\gamma|^2+|\delta|^2(1-\eta_{\text c})
\bigr).
\label{eq:P_mixed_app}
\end{align}
Accepting all four fixed-photon-number patterns gives
\begin{equation}
P^{\Psi}_{\text{tot}}=4P^{\Psi}_{\text{succ}}.
\label{eq:Ppsi_total_app}
\end{equation}

The probability of obtaining any of the six accepted two-click patterns is
\begin{widetext}
\begin{align}
R^{\text{tot}}_{\text{succ}}
&=2P^{\Phi}_{\text{succ}}+4P^{\Psi}_{\text{succ}}
\nonumber\\
&=\frac{\eta_{\text c}^2}{2}\Bigl[
|\beta|^4\bigl(
|\gamma|^2+|\delta|^2(1-\eta_{\text c})\bigr)^2
+
|\delta|^4\bigl(|\alpha|^2+|\beta|^2(1-\eta_{\text c})\bigr)^2
+4|\beta|^2|\delta|^2
\bigl(|\alpha|^2+|\beta|^2(1-\eta_{\text c})
\bigr)
\bigl(|\gamma|^2+|\delta|^2(1-\eta_{\text c})
\bigr)
\Bigr].
\label{eq:P_any_app}
\end{align}
\end{widetext}
This expression includes both genuine and loss-induced false-heralding events.

For the symmetric family, all six patterns occur with the same probability:
\begin{equation}
P^{\Phi}_{\text{succ}}
=P^{\Psi}_{\text{succ}}
=\frac{\eta_{\text c}^2}{2}(1-q)^2
\left[q+(1-q)(1-\eta_{\text c})
\right]^2.
\label{eq:equal_pattern_probabilities_app}
\end{equation}
Consequently,
\begin{equation}
P^{\Phi}_{\text{tot}}
=\eta_{\text c}^2(1-q)^2
\left[q+(1-q)(1-\eta_{\text c})
\right]^2,
\qquad
P^{\Psi}_{\text{tot}}
=2P^{\Phi}_{\text{tot}},
\label{eq:class_probabilities_app}
\end{equation}
and
\begin{equation}
R^{\text{tot}}_{\text{succ}}
=3\eta_{\text c}^2(1-q)^2
\left[q+(1-q)(1-\eta_{\text c})
\right]^2.
\label{eq:R_family_app}
\end{equation}
All heralding probabilities therefore retain the quadratic scaling
$O(\eta_{\text c}^2)$. The joint probability that either target-state pattern is produced by the genuine coherent component is
\begin{equation}
R_{\text{true}}^\Phi
=\eta_{\text c}^2q^2(1-q)^2.
\label{eq:R_true_family_app}
\end{equation}

For a fixed-photon-number pattern such as $(D_1,D_3)$, the two loss-free contributions superpose coherently into
\begin{equation}
\ket{\widetilde{\Phi}_{13}}
=\alpha\beta\gamma\delta
\left(
\ket{0110}-\ket{1001}
\right).
\label{eq:mixed_state_app}
\end{equation}
Whenever its heralding probability is nonzero, this component is maximally entangled independently of the source parameters. The loss-induced contributions populate states orthogonal to this target subspace and enter the conditional state incoherently after the environment is traced out.

The fidelity with the corresponding normalized target state
$\ket{\Psi}=(\ket{0110}-\ket{1001})/\sqrt{2}$ is
\begin{equation}
\mathcal F_{\Psi}
=\frac{
|\alpha|^2|\gamma|^2}{
\bigl(|\alpha|^2+|\beta|^2(1-\eta_{\text c})
\bigr)
\bigl(|\gamma|^2+|\delta|^2(1-\eta_{\text c})
\bigr)}.
\label{eq:F_mixed_app}
\end{equation}
For the symmetric family, this reduces to
\begin{equation}
\mathcal F_{\Psi}=\mathcal F_{\Phi}
=\left[\frac{q}{q+(1-q)(1-\eta_{\text c})}
\right]^2.
\label{eq:F_family_app}
\end{equation}
Thus, under the symmetric condition, all six patterns have the same per-pattern heralding probability and the same fidelity with their respective ideal target states. The vacuum--four-photon and fixed-two-photon outputs nevertheless remain physically distinct photon-number encodings.

%=======================

\section{Key Rate Simulation}\label{app:keyratesimulation}
We describe the simulation model used to calculate the DI-CKA key rates in Sec.~\ref{sec:long-distanceDI-CKA}. In particular, we derive the conditional quantum state heralded among the four parties. The key rate can be calculated from a probability distribution of measurement results when the parties perform the measurements described in Sec.~\ref{sec:long-distanceDI-CKA} on the distributed quantum state.

We model channel loss and detection efficiency of the parties using beam splitters with transmissivities $\eta_c$ and $\eta$, respectively. Applying the beam-splitter transformation to the Type-A source state gives 
\begin{equation}
    \begin{split}
        &\alpha \ket{10}_{a_ia_i'} \ket{0}_{E_{a_i}} \ket{0}_{F_{a_i}} + \beta \ket{01}_{a_ia_i'} \ket{0}_{E_{a_i}} \ket{0}_{F_{a_i}}\\
        \rightarrow&\beta \sqrt{\eta_c} \ket{000}_{a_iE_{a_i}F_{a_i}} \ket{1}_{a_i'} + (\alpha \sqrt{\eta} \ket{100}\\
        +& \alpha\sqrt{1-\eta} \ket{001} + \beta\sqrt{1-\eta_c} \ket{010})_{a_iE_{a_i}F_{a_i}} \ket{0}_{a_i'}\\
        \equiv  &\ket{S_A}_{a_iE_{a_i}F_{a_i}} \ket{1}_{a_i'} + \ket{V_A}_{a_iE_{a_i}F_{a_i}} \ket{0}_{a_i'},
    \end{split}
\end{equation}
where we define
\begin{align}
    \ket{S_A} &= \beta \sqrt{\eta_c} \ket{000},\\
    \ket{V_A} &= \alpha( \sqrt{\eta} \ket{100} + \sqrt{1-\eta} \ket{001}) \notag \\
    &+ \beta \sqrt{1-\eta_c} \ket{010}.
\end{align}
Here, $\ket{S_A}$ and $\ket{V_A}$ are the unnormalized states of the retained and environment modes associated with one photon and vacuum, respectively in the transmitted mode. $E_{a_i}$ and $F_{a_i}$ represent environmental modes. Similarly, the Type-B state transforms as follows
\begin{widetext}
\begin{equation}
    \begin{split}
        &\gamma \ket{00}_{b_ib_i'} \ket{0}_{E_{b_i}} \ket{0}_{F_{b_i}} + \delta \ket{11}_{b_ib_i'} \ket{0}_{E_{b_i}} \ket{0}_{F_{b_i}}\\
        \rightarrow&\delta \sqrt{\eta_c} (\sqrt{\eta} \ket{100} + \sqrt{1-\eta} \ket{001})_{b_iE_{b_i}F_{b_i}} \ket{1}_{b_i'} 
        + (\gamma \ket{000} + \delta \sqrt{1-\eta_c} (\sqrt{\eta} \ket{110} + \sqrt{1-\eta} \ket{011}))_{b_iE_{b_i}F_{b_i}}\ket{0}_{b_i'}\\
        \equiv  &\ket{S_B}_{b_iE_{b_i}F_{b_i}} \ket{1}_{b_i'} + \ket{V_B}_{b_iE_{b_i}F_{b_i}} \ket{0}_{b_i'},
    \end{split}
\end{equation}
\end{widetext}
where
\begin{align}
    \ket{S_B} &= \delta \sqrt{\eta_c} (\sqrt{\eta} \ket{100} + \sqrt{1-\eta} \ket{001}),\\
    \ket{V_B} &= \gamma \ket{000} \notag\\
    &+ \delta \sqrt{1-\eta_c} (\sqrt{\eta} \ket{110} + \sqrt{1-\eta} \ket{011}),
\end{align}
and $E_{b_i}$ and $F_{b_i}$ represent environmental modes. Measurement operators of the PNR detectors at the central station with dark count probability $p_\text{DC}$ can be expressed as
\begin{equation}
    \{ (1-p_\text{DC}) \ketbra{0}, \ketbra{1} + p_\text{DC} \ketbra{0}  \}.
\end{equation}
Here, we discard clicks corresponding to more than one photon. When only the detectors $D_1$ and $D_2$ click, the total measurement operator of all the detectors is
\begin{equation}\label{eq:totalmeasurementoperator}
    \begin{split}
        &\left( \ketbra{1}{1} + p_\text{DC} \ketbra{0}{0} \right)_{D_1} \left( \ketbra{1}{1} + p_\text{DC} \ketbra{0}{0} \right)_{D_2}\\
        \otimes &\left( (1-p_\text{DC}) \ketbra{0}{0} \right)_{D_3}\left( (1-p_\text{DC}) \ketbra{0}{0} \right)_{D_4}\\
        = &\left(1-p_\text{DC}\right)^2 \ketbra{1100}{1100}\\
        + &\left(1-p_\text{DC}\right)^2p_\text{DC} \ketbra{1000}{1000}\\
        + &\left(1-p_\text{DC}\right)^2 p_\text{DC} \ketbra{0100}{0100}\\
        + & \left( 1-p_\text{DC} \right)^2 p_\text{DC}^2 \ketbra{0000}{0000}.
    \end{split}
\end{equation}
From the operation of the beam-splitter network at the central station shown in Eq.~(\ref{eq:U_matrix_appendix}), we can calculate a quantum state entering the interferometer corresponding to each term in Eq.~(\ref{eq:totalmeasurementoperator}). A quantum state corresponding to the first term can be expressed as
\begin{equation}
    \ket{\phi^1} = \frac{1}{2} \left( \ket{S_A} \ket{V_B} \ket{S_A} \ket{V_B} - \ket{V_A} \ket{S_B} \ket{V_A} \ket{S_B} \right).
\end{equation}
Quantum states corresponding to the second and the third terms are, respectively,
\begin{equation}
\begin{split}
    \ket{\phi^2} =  \frac{1}{2} ( &\ket{S_A}\ket{V_B}\ket{V_A}\ket{V_B} + \ket{V_A}\ket{S_B}\ket{V_A}\ket{V_B}\\
    + &\ket{V_A}\ket{V_B}\ket{S_A}\ket{V_B} + \ket{V_A}\ket{V_B}\ket{V_A}\ket{S_B} ),
\end{split}
\end{equation}
\begin{equation}
\begin{split}
    \ket{\phi^3} = \frac{1}{2} ( &\ket{S_A}\ket{V_B}\ket{V_A}\ket{V_B} - \ket{V_A}\ket{S_B}\ket{V_A}\ket{V_B} \\
    + &\ket{V_A}\ket{V_B}\ket{S_A}\ket{V_B} - \ket{V_A}\ket{V_B}\ket{V_A}\ket{S_B} ).
\end{split}
\end{equation}
Finally, a quantum state corresponding to the last term is
\begin{equation}
    \ket{\phi^4} = \ket{V_A}\ket{V_B}\ket{V_A}\ket{V_B}.
\end{equation}
Then, the conditional joint state of the retained and environmental modes is
\begin{equation}
\begin{split}
    \rho_\text{joint} = \frac{1}{P_\text{dist}} ( &P_1 \ketbra{\phi^1}{\phi^1}+ P_2 \ketbra{\phi^2}{\phi^2}  \\
    + &P_3 \ketbra{\phi^3}{\phi^3} + P_4 \ketbra{\phi^4}{\phi^4} ),
\end{split}
\end{equation}
where
\begin{equation}
    \begin{split}
        P_1 &= (1-p_{DC} )^2,\\
        P_2 &= P_3 = (1-p_{DC})^2 p_{DC},\\
        P_4 &= (1-p_{DC})^2 p_{DC}^2,
    \end{split}
\end{equation}
and
\begin{equation}
    \begin{split}
        P_\text{dist} = \text{Tr} [&P_1 \ketbra{\phi^1}{\phi^1} + P_2 \ketbra{\phi^2}{\phi^2}\\
        + &P_3 \ketbra{\phi^3}{\phi^3} + P_4 \ketbra{\phi^4}{\phi^4}].
    \end{split}
\end{equation}
Tracing out the environmental modes gives the state shared by the four parties, that is,
\begin{equation}
    \rho_\text{parties} = \text{Tr}_{E,F} [\rho_\text{joint}].
\end{equation}
In addition, the successful heralding probability is $P_\text{herald} = P_\text{dist}$.
%=======================
\section{Performance with SPDC Sources}
\label{app:performance_with_spdc}

Here, we replace the two ideal Type-B resources with two-mode squeezed-vacuum (TMSV) states generated by spontaneous parametric down-conversion (SPDC). We work in the low-gain regime and truncate each source at two emitted pairs, thereby retaining the leading multipair contribution within a finite-dimensional model.

\subsection{Heralded State with Truncated TMSV Sources}

To isolate the effect of multipair emission, we initially assume lossless channels, an ideal central interferometer, and ideal PNR detection.

The truncated Type-B state is
\begin{equation}
\begin{aligned}
\ket{\phi_i}_{b_i b_i'}
&=\frac{\ket{0,0}
+\lambda_i\ket{1,1}
+\lambda_i^2\ket{2,2}}{
\sqrt{Z_i}},
\\
Z_i&=1+|\lambda_i|^2+|\lambda_i|^4,
\end{aligned}
\label{eq:truncated-tmsv-B-source}
\end{equation}
where $|\lambda_i|<1$ is proportional to the TMSV gain. The truncation in Eq.~\eqref{eq:truncated-tmsv-B-source} is a modelling approximation motivated by the following. In the lossless case, with PNRDs at the central station, the accepted heralding event contains exactly two photons. Since the central interferometer preserves the total photon number, components containing more than two pairs have zero overlap with the accepted event. Retaining up to the two-pair component therefore yields the exact normalized conditional state under these assumptions. In the lossy case, higher-order emissions may also produce accepted events if the additional photons are lost. In this case, the truncation is justified by operating in the weak-squeezing regime. Specifically, we take $\lambda_i\leq0.1$, for which the probability that each Type-B source emits three or more pairs is bounded by $P_i(n\geq3)=|\lambda_i|^6\leq 10^{-6}$. Thus, truncating the TMSV state at the two-pair level provides a sufficiently accurate description in the weak-squeezing regime considered here.

The global input state is
\begin{equation}
\ket{\Psi_{\text{in}}}
=\ket{\psi_1}_{a_1a_1'}
\otimes
\ket{\phi_1}_{b_1b_1'}
\otimes
\ket{\psi_2}_{a_2a_2'}
\otimes
\ket{\phi_2}_{b_2b_2'},
\label{eq:global_input_app2}
\end{equation}
with the Type-A states $\ket{\psi_i}$ defined in \Cref{eq:typeA}. We use the mode ordering $(a_1,b_1,a_2,b_2)$ for the retained modes and $(a_1',b_1',a_2',b_2')$ for the transmitted modes. The complete expansion of Eq.~(\ref{eq:global_input_app2}) contains $2\times3\times2\times3=36$ terms. Its complete expansion  is
\begin{widetext}
\begin{align}
\ket{\Psi_{\text{in}}}
={}\mathcal N_B&\Bigg\{
\alpha_1\alpha_2\Big[
      \ket{1010}\ket{0000}
+\lambda_2\ket{1011}\ket{0001}
+\lambda_2^2\ket{1012}\ket{0002}
+\lambda_1\ket{1110}\ket{0100}
+\lambda_1\lambda_2\ket{1111}\ket{0101}\nonumber\\&\hspace{1mm}+\lambda_1\lambda_2^2\ket{1112}\ket{0102}
+\lambda_1^2\ket{1210}\ket{0200}
+\lambda_1^2\lambda_2\ket{1211}\ket{0201}
+\lambda_1^2\lambda_2^2\ket{1212}\ket{0202}
\Big]
\nonumber\\&\hspace{1mm}+\alpha_1\beta_2\Big[
      \ket{1000}\ket{0010}
+\lambda_2\ket{1001}\ket{0011}
+\lambda_2^2\ket{1002}\ket{0012}
+\lambda_1\ket{1100}\ket{0110}
+\lambda_1\lambda_2\ket{1101}\ket{0111}\nonumber\\&\hspace{1mm}
+\lambda_1\lambda_2^2\ket{1102}\ket{0112}
+\lambda_1^2\ket{1200}\ket{0210}
+\lambda_1^2\lambda_2\ket{1201}\ket{0211}
+\lambda_1^2\lambda_2^2\ket{1202}\ket{0212}
\Big]
\nonumber\\[1mm]
&+\beta_1\alpha_2\Big[
      \ket{0010}\ket{1000}
+\lambda_2\ket{0011}\ket{1001}
+\lambda_2^2\ket{0012}\ket{1002}
+\lambda_1\ket{0110}\ket{1100}
+\lambda_1\lambda_2\ket{0111}\ket{1101}\nonumber\\&\hspace{1mm}
+\lambda_1\lambda_2^2\ket{0112}\ket{1102}
+\lambda_1^2\ket{0210}\ket{1200}
+\lambda_1^2\lambda_2\ket{0211}\ket{1201}
+\lambda_1^2\lambda_2^2\ket{0212}\ket{1202}
\Big]
\nonumber\\[1mm]
&+\beta_1\beta_2\Big[
      \ket{0000}\ket{1010}
+\lambda_2\ket{0001}\ket{1011}
+\lambda_2^2\ket{0002}\ket{1012}
+\lambda_1\ket{0100}\ket{1110}
+\lambda_1\lambda_2\ket{0101}\ket{1111}\nonumber\\&\hspace{1mm}
+\lambda_1\lambda_2^2\ket{0102}\ket{1112}
+\lambda_1^2\ket{0200}\ket{1210}
+\lambda_1^2\lambda_2\ket{0201}\ket{1211}
+\lambda_1^2\lambda_2^2\ket{0202}\ket{1212}
\Big]\Bigg\},
\label{eq:tmsv-global-input-full}
\end{align}
\end{widetext}
where $\mathcal N_B=1/\sqrt{Z_1Z_2}$.

The projectors derived in Appendix~\ref{app:projectors} under the restriction to the local $\{\ket{0},\ket{1}\}$ subspace cannot be used unchanged once multipair emission is included. In particular, two photons may now enter the interferometer through a single Type-B mode, so the complete inverse image of the detected output state must be retained within the two-photon input sector.

For clicks at $D_1$ and $D_2$, the detected output state is $\ket{1100}_d$. Applying the inverse interferometer transformation gives
\begin{align}
\ket{1100}_{d}
&\xrightarrow{\hat{\mathcal U}^{\dagger}}
\ket{\chi_{12}}
\nonumber\\
&=\frac{1}{4}\Big(2\ket{1010}-2\ket{0101}
\nonumber\\[-1mm]
&\hspace{8mm}
+\sqrt{2}\ket{2000}-\sqrt{2}\ket{0200}
\nonumber\\[-1mm]
&\hspace{8mm}
+\sqrt{2}\ket{0020}-\sqrt{2}\ket{0002}
\Big).
\label{eq:full_proj12}
\end{align}

The first two terms reproduce the projector obtained under the $0$--$1$-photon restriction.  The remaining four terms are bunched two-photon inputs and are essential once multipair emission is allowed.

The occupations $\ket{2000}$ and $\ket{0020}$ do not occur in Eq.~(\ref{eq:global_input_app2}), because each Type-A source transmits at most one photon. Projecting the global input state onto Eq.~(\ref{eq:full_proj12}) therefore gives the unnormalized retained state
\begin{align}
\ket{\widetilde{\Phi}_{12}}=
\frac{\mathcal N_B}{2}\Bigg[
&\beta_1\beta_2\ket{0000}-
\alpha_1\alpha_2\lambda_1\lambda_2\ket{1111}
\nonumber\\[-1mm]
&-\frac{\alpha_1\alpha_2}{\sqrt{2}}
\left(\lambda_1^2\ket{1210}
+\lambda_2^2\ket{1012}
\right)
\Bigg].
\label{eq:truncated-tmsv-heralded-unnormalized}
\end{align}
The first two terms form the desired vacuum--four-photon coherence. The states $\ket{1210}$ and $\ket{1012}$ arise when one Type-B source emits two pairs while the other emits vacuum. These states lie outside the local qubit subspace and constitute multipair leakage.

Defining
\begin{align}
D_{12}
={}&
|\beta_1\beta_2|^2
+|\alpha_1\alpha_2\lambda_1\lambda_2|^2
\nonumber\\
&+\frac{|\alpha_1\alpha_2|^2}{2}
\left(|\lambda_1|^4+|\lambda_2|^4
\right),
\label{eq:truncated-tmsv-D12}
\end{align}
the probability of this detection pattern is
\begin{equation}
P_{12}=\frac{\mathcal N_B^2}{4}D_{12},
\label{eq:truncated-tmsv-P12}
\end{equation}
and the normalized conditional state is
\begin{align}
\ket{\Phi_{12}^{\text{SPDC}}}=
\frac{1}{\sqrt{D_{12}}}\Bigg[
&\beta_1\beta_2\ket{0000}
-\alpha_1\alpha_2\lambda_1\lambda_2\ket{1111}
\nonumber\\[-1mm]
&-\frac{\alpha_1\alpha_2}{\sqrt{2}}
\left(\lambda_1^2\ket{1210}
+\lambda_2^2\ket{1012}\right)
\Bigg].
\label{eq:truncated-tmsv-heralded-normalized}
\end{align}

We now assume identical TMSV parameters, $\lambda_1=\lambda_2\equiv \lambda$, and align the relevant source phases. In the phase convention associated with this detection pattern, the target state is
\begin{equation}
\ket{\Phi_4(q)}
=\sqrt{1-q}\ket{0000}-\sqrt{q}\ket{1111}.
\label{eq:truncated-tmsv-target}
\end{equation}
Optimizing the overlap with this target over the normalized Type-A source amplitudes gives the condition
\begin{equation}
\frac{\alpha_1\alpha_2\lambda^2}
{\beta_1\beta_2}=
\sqrt{\frac{q}{4(1-q)}},
\label{eq:truncated-tmsv-balance-condition}
\end{equation}
with the phases chosen consistently. This condition can always be satisfied by an appropriate choice of normalized Type-A source amplitudes for $0<q<1$.

Under Eq.~(\ref{eq:truncated-tmsv-balance-condition}), the normalized conditional state becomes
\begin{align}
\ket{\Phi_{12}^{\text{SPDC}}(q)}
=\frac{1}{\sqrt{2(2-q)}}\Bigg[
&2\sqrt{1-q}\ket{0000}
-\sqrt{q}\ket{1111}
\nonumber\\[-1mm]
&-\sqrt{\frac{q}{2}}
\left(\ket{1210}+\ket{1012}
\right)
\Bigg].
\label{eq:truncated-tmsv-q-state}
\end{align}
Its maximum lossless fidelity with the target state is therefore
\begin{equation}
\mathcal F_{\text{SPDC}}^{\max}(q)
=\max_{\{\alpha_i,\beta_i\}}
\left|\braket{\Phi_4(q)}{\Phi_{12}^{\text{SPDC}}}\right|^2
=\frac{2-q}{2}.
\label{eq:optimized_spdc_fidelity}
\end{equation}
For a maximally entangled target, $q=1/2$, the maximum fidelity is therefore $3/4$. As $q\to0$, the target becomes increasingly vacuum dominated and the maximum fidelity approaches unity, at the cost of diminishing entanglement. 

Thus, replacing the ideal Type-B sources with TMSV sources prevents the central measurement from heralding a pure GHZ state in the intended single-rail qubit subspace. Double-pair emission produces leakage at the same order in the squeezing parameter as the desired $\ket{1111}$ contribution, so reducing the SPDC gain does not eliminate this relative contamination. Photon loss opens additional false-heralding pathways by allowing higher-photon-number inputs to produce the accepted two-photon detection pattern. Nevertheless, a vacuum-dominated nonmaximally entangled target can be heralded with high fidelity because both the desired $\ket{1111}$ component and the leading double-pair leakage become small relative to the vacuum component. 

\subsection{Parity--CHSH Test with Displacement-Based Measurements}
\label{subsec:truncated-tmsv-displacement-pchsh}

Since vacuum-biased GHZ states can exhibit enhanced loss tolerance, we next investigate whether the full realistically heralded state, including multipair leakage, can violate the parity--CHSH inequality using displacement-based measurements. After removing an overall phase and amplitude, the state can be written in the physical mode ordering $(a_1,b_1,a_2,b_2)$ as
\begin{align}
\ket{\psi(k,\kappa)}
=\frac{1}{\sqrt{\mathcal D}}\Bigg[
&\ket{0000}-k\ket{1111}
\nonumber\\[-1mm]
&-\frac{k}{\sqrt{2}}\left(\kappa\ket{1210}
+\kappa^{-1}\ket{1012}\right)\Bigg],
\label{eq:pchsh-disp-truncated-state}
\end{align}
where
\begin{align}
k=\left|\frac{\alpha_1\alpha_2\lambda_1\lambda_2}{
\beta_1\beta_2}\right|,
\kappa=\left|\frac{\lambda_1}{\lambda_2}\right|,
\mathcal D=
1+k^2\left[1+\frac{\kappa^2+\kappa^{-2}}{2}\right].
\label{eq:pchsh-disp-state-parameters}
\end{align}
Thus, $\kappa=1$ corresponds to identical Type-B TMSV sources, whereas $\kappa\neq1$ describes unequal pair-production parameters.

For each party, the measurement consists of a displacement $\hat D(\xi)$ followed by an on--off detector with efficiency $\eta$. We can modify the displacement-based measurement explained in Sec.~ \ref{subsec:displacement} in the main text to incorporate the detection efficiency in the measurement operator.  In the absence of dark counts, the no-click and click POVM elements are
\begin{equation}
\hat\Pi_{0}^{(\eta)}=(1-\eta)^{\hat n},
\qquad
\hat\Pi_{1}^{(\eta)}
=I-\hat\Pi_{0}^{(\eta)},
\label{eq:onoff-povm}
\end{equation}
respectively. Assigning no click to $+1$ and click to $-1$, the effective observable acting on the state before the displacement is
\begin{align}
\hat M^{(\eta)}(\xi)&=\hat D^\dagger(\xi)
\left(\hat\Pi_{0}^{(\eta)}-
\hat\Pi_{1}^{(\eta)}\right)
\hat D(\xi)
\nonumber\\
&=2\hat D^\dagger(\xi)
(1-\eta)^{\hat n}
\hat D(\xi)-I.
\label{eq:lossy-displacement-observable}
\end{align}
All click and no-click events are retained, so the Bell test does not involve detection postselection.

For four parties, the parity--CHSH expression is
\begin{align}
S_4
={}&\langle A_0B_0\rangle
+\langle A_0B_1\rangle\nonumber\\
&+\langle A_1B_0C_1^{(1)}C_1^{(2)}\rangle-\langle A_1B_1C_1^{(1)}C^{(2)}_1\rangle,
\label{eq:pchsh-disp-score}
\end{align}
with the local bound $S_4\leq2$.

As in Sec.~\ref{sec:2}, the parity--CHSH expression distinguishes the two-setting roles $A$ and $B$ from the remaining parties, so their assignment to physical modes can affect the violation. For the state in Eq.~\eqref{eq:pchsh-disp-truncated-state}, the Type-A modes $a_1$ and $a_2$ have equal photon occupations in every component. The Type-B modes $b_1$ and $b_2$, by contrast, have complementary occupations, $20$ and $02$, in the two multipair components. We therefore include the assignment of modes to measurement roles in the optimization.

\begin{figure}[t!]
    \centering
    \includegraphics[width=\linewidth]{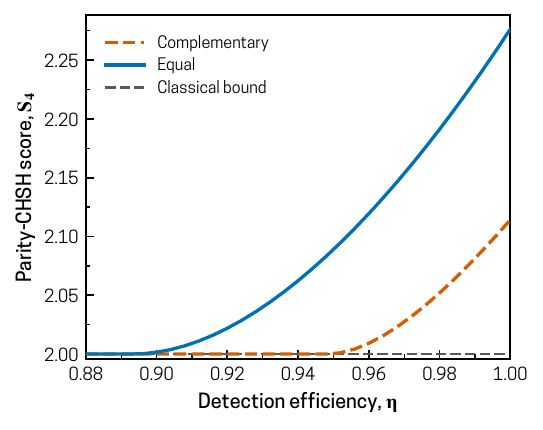}
    \caption{Optimized parity--CHSH value $S_4$ as a function of the common local detection efficiency $\eta$ for the heralded state obtained with SPDC-based Type-B sources. The solid blue curve (Equal) assigns the two-setting roles to the two Type-A parties, $\mathsf{A}_1$ and $\mathsf{A}_2$, whereas the dashed orange curve (Complementary) assigns them to one Type-A and one Type-B party, $\mathsf{A}_1$ and $\mathsf{B}_1$; the remaining parties perform one measurement setting. All source parameters and displacement amplitudes are optimized independently at each efficiency. The horizontal dashed gray line denotes the classical bound $S_4=2$.}
    \label{fig:PCHS_with_TMSV}
\end{figure}

After jointly optimizing the source parameters and displacement amplitudes at each detection efficiency, we obtain the parity--CHSH values shown in Fig.~\ref{fig:PCHS_with_TMSV}. When one Type-A and one Type-B party are assigned the two-setting roles, the detection-efficiency threshold is $\eta_{\mathrm{th}}=0.9487$. Assigning both two-setting roles to the Type-A parties lowers this threshold to $\eta_{\mathrm{th}}=0.8932$ and yields a much larger violation throughout the relevant efficiency range. The separation between the two curves demonstrates that the assignment of physical parties to measurement roles is a substantive network-design parameter rather than a mere relabeling convention. 

%================

\bibliography{apssamp}

\end{document}